\documentclass[final,onefignum,onetabnum]{siamonline220329}

\usepackage{comment}

\usepackage{lipsum}
\usepackage{amsfonts}
\usepackage{graphicx}
\usepackage{epstopdf}
\usepackage{algorithmic}
\ifpdf
  \DeclareGraphicsExtensions{.eps,.pdf,.png,.jpg}
\else
  \DeclareGraphicsExtensions{.eps}
\fi

\usepackage{enumitem}
\setlist[enumerate]{leftmargin=.5in}
\setlist[itemize]{leftmargin=.5in}

\newsiamremark{remark}{Remark}
\newsiamremark{hypothesis}{Hypothesis}
\crefname{hypothesis}{Hypothesis}{Hypotheses}
\newsiamthm{claim}{Claim}

\headers{Dynamical mean field theory for SGD}{C. Gerbelot, E. Troiani, F. Mignacco, F. Krzakala, L. Zdeborova}

\title{Rigorous dynamical mean field theory
for stochastic gradient descent methods \thanks{
\funding{This work was funded by the ERC under the European Union’s Horizon 2020 Research and Innovation Program Grant Agreement 714608-SMiLe, as well as by the Swiss National Science Foundation grant SNFS OperaGOST, $200021\_200390$.}
}}

\author{C\'edric Gerbelot\thanks{Laboratoire de Physique de l'Ecole Normale Supérieure, Université PSL, Paris, France and Courant Institute of Mathematical Sciences, New York University, New York, NY 10012, USA}
\and Emanuele Troiani\thanks{Statistical Physics of Computation Lab, \'Ecole Polytechnique F\'ed\'erale de Lausanne (EPFL).}
\and Francesca Mignacco\thanks{Université Paris-Saclay, CNRS, CEA, Institut de physique théorique, 91191, Gif-sur-Yvette, France.}
\and Florent Krzakala\thanks{Information, Learning and Physics lab, \'Ecole Polytechnique F\'ed\'erale de Lausanne (EPFL).}
\and Lenka Zdeborov\'a\footnotemark[3]
}

\usepackage{amsopn}

\def\max{{\rm max}}

\def\XX{{\bm{X}}}
\def\yy{{\bm{y}}}
\def\ww{{\bm{w}}}
\def\vv{{\bm{v}}}

\def\wwt{{\bm{w^*}}}
\def\rr{{\bm{r}}}

\def\reg{\mathbf{F}}

\newcommand{\approxP}{\mathrel{\stackrel{{\rm P}}{\mathrel{\scalebox{1.8}[1]{$\simeq$}}}}}
\usepackage{physics}
\usepackage{amsmath}
\usepackage{amssymb}
\usepackage{bm}
\newcommand{\bZ}{{\mathbf{Z}}}
\newcommand{\bU}{{\mathbf{U}}}
\newcommand{\bV}{{\mathbf{V}}}
\newcommand{\cS}{\mathcal{S}}

\newcommand{\bGamma}{{\boldsymbol{\Gamma}}}
\newcommand{\btheta}{{\boldsymbol{\theta}}}
\newcommand{\bTheta}{{\boldsymbol{\Theta}}}
\newcommand{\beeta}{{\boldsymbol{\eta}}}
\newcommand{\bomega}{{\boldsymbol{\omega}}}

\newcommand{\bnu}{{\boldsymbol{\nu}}}
\renewcommand{\mathbf}{\bm}
\renewcommand{\ell}{l}
\ifpdf
\hypersetup{
  pdftitle={An Example Article},
  pdfauthor={D. Doe, P. T. Frank, and J. E. Smith}
}
\fi

\begin{document}

\maketitle

\begin{abstract}
    We prove closed-form equations for the exact high-dimensional asymptotics of a family of first order gradient-based methods, learning an estimator (e.g. M-estimator, shallow neural network, ...)
    from observations on Gaussian data with empirical risk minimization. This includes widely used algorithms such as stochastic gradient descent (SGD) or Nesterov acceleration. The obtained equations match those resulting from the discretization of dynamical mean-field theory (DMFT) equations from statistical physics when 
    applied to the corresponding gradient flow. Our proof method allows us to 
    give an explicit description of how memory kernels build up in the effective dynamics, and to include non-separable update functions, allowing datasets with non-identity covariance matrices. Finally, we provide numerical implementations of the equations for SGD with generic extensive batch-size and constant learning rates.
    \end{abstract}



\section{Introduction}
        Stochastic gradient descent methods are one of the cornerstones of optimization and thus, modern machine-learning. Notably, stochastic gradient descent and its variants have become the 
        method of choice for the optimization of large deep learning architectures, see e.g. \cite{lecun1998gradient,kingma2014adam,rumelhart1986learning}. However, gradient based dynamics are not 
        restricted to the field of machine learning and computational mathematics, as they are also at the center of out-of-equilibrium 
        statistical mechanics through the notion of Langevin dynamics, see e.g. \cite{mezard1987spin}. Obtaining an exact understanding of these procedures has been a long-standing 
        problem, notably for disordered systems, e.g. spin glasses, where a significant set of results has been obtained, first using heuristic, theoretical physics methods \cite{sompolinsky1981dynamic,sompolinsky1982relaxational,crisanti1993sphericalp,cugliandolo1993analytical} and then rigorous probability theory \cite{arous2001aging,ben2006cugliandolo,celentano2021high,liang2022high}. 
        In theoretical physics, the effective dynamics describing the high-dimensional behavior of gradient flow is called dynamical mean-field theory (DMFT), in reference to the reduction of a system of strongly correlated degrees of freedom to low-dimensional order parameters whose evolution can be tracked analytically by a set of self-consistent equations. In the continuous time limit, those equations take the form of a stochastic integro-differential system involving memory kernels and additive Gaussian processes, whose parameters are entirely characterized by the parameters of the system, such as the form of the gradient or the temperature of the thermal noise. In recent years, DMFT equations have been used by physicists to study a wide variety of high-dimensional disordered dynamical systems (see, e.g., \cite{maimbourg2016solution,szamel2017simple,MSZ20,Roy_2019}), including dynamical aspects of constraint satisfaction problems and gradient descent methods in the context of statistical learning \cite{agoritsas2018out,  mannelli2021,sclocchi2022high}. In particular, \cite{mignacco2020dynamical,mignacco2021stochasticity,Mignacco_2022} have applied DMFT equations to analyze the high-dimensional dynamics of the SGD algorithm by modelling the mini-batch sampling via selection variables obeying an independent pointwise stochastic process.
        
        While the recent work of \cite{celentano2021high} provides game-changing progress into the rigorous establishment of the DMFT, it does not account for the stochasticity of the gradient descent algorithms and their proof is limited to the data matrix to be random, with i.i.d. centered subgaussian entries. In the present work we remove these two limitations and establish the DMFT equations for a broad class of stochastic algorithms (including SGD, various momentum methods or Langevin algorithms), and for a broader class of data (including Gaussian with a rather generic covariance). 
        
        Theoretical physics works on DMFT aim to describe the continuous time dynamics, because the physical dynamics simply is continuous. When gradient based methods are used as algorithms they are always run in discrete time and thus for algorithmic purposes the analysis of the discrete dynamics is of larger interest in data science. In previous theoretical physics works the derivation of DMFT equations is always presented for the continuous (flow) limit of the dynamics.
        In this paper we prove that the discrete DMFT equations provide exact asymptotic analysis for the discrete gradient descent methods as well. This has been already showed empirically in \cite{mignacco2020dynamical,mignacco2021stochasticity} and the discrete-time -- albeit non rigorous -- equations have been applied in \cite{Mignacco_2022} to study the impact of a discrete time step on SGD noise. While a larger part of \cite{celentano2021high} is devoted to proving the continuous-time equations, they also establish the discrete time DMFT. In the present paper we will only consider the discrete version because (a) our main motivation is analysis of actual algorithms, (b) the exactness of the discrete DMFT is not discussed in the literature and we thus want to rectify that.

        Our proof of dynamical mean-field theory equations applies to a wide range of supervised learning problems, where an estimator is learned using stochastic gradient descent on a cost function defined by empirical risk minimization. In this regard, consider the following optimization problem
        \begin{align}
        &\hat{{\ww}} \in \inf_{\ww \in \mathbb{R}^{d \times q}} \mathcal{L}(\XX\ww,\mathbf{y})+\reg(\ww) \\
        &\mbox{where} \thickspace \yy = \Phi_{0}\left(\XX\wwt\right),
        \label{eq:ERM_definition}
    \end{align}
    where $\XX \in \mathbb{R}^{n \times d}$ is the design matrix, the observed labels $\yy \in \mathbb{R}^{n}$ are generated according to a ground truth parametrized by a continuous, separable function $\Phi_{0} : \mathbb{R}^{n\times q} \to \mathbb{R}^n$ and ground-truth vector $\wwt \in \mathbb{R}^{d \times q}$, and the loss and regularization $\mathcal{L} : \mathbb{R}^{n \times q}\times \mathbb{R}^{n} \to \mathbb{R},\reg : \mathbb{R}^{d \times q} \to \mathbb{R}$ are differentiable functions. A function $\mathbf{F} : \mathbb{R}^{d} \to \mathbb{R}^{d}$ is said to be separable if it is defined by 
    a scalar function $f : \mathbb{R} \to \mathbb{R}$ applied to each element of the input value, i.e. 
    \begin{equation}
        \mathbf{F}(\mathbf{x}) = \begin{bmatrix}f(x_{1}) \\f(x_{2}) \\ ... \\ f(x_{d})\end{bmatrix}.
    \end{equation}
     In the case where $\mathbf{F}$ is scalar valued, the definition of separable becomes 
    \begin{equation}
        \mathbf{F}(\mathbf{x}) = \sum_{i=1}^{d}f(x_{i}).
    \end{equation}  The number of samples $n$ and dimension of the inputs $d$ will be taken to infinity (the high-dimensional limit), while the number of weight vectors $q$ (corresponding to the number of hidden units) will remain finite.
    We will consider a generic family of  discrete-time dynamics in Theorem \ref{th:main_dmft}, which includes stochastic gradient descent methods widely used in practice: 
    a candidate $\hat{\ww}$ is estimated using gradient descent by producing the following sequence of iterates 
    \begin{align}
    \label{eq:grad_desc}
        \ww^{t+1} = \ww^{t}-\gamma^{t}\left(\XX^{\top}\nabla \mathcal{L}^{t}(\XX\ww^{t},\yy)+\nabla \reg(\ww^{t})\right)
    \end{align}
    where $\gamma^{t}$ is the scalar learning rate, $\nabla \mathcal{L}^{t}(.,\mathbf{y}) \in \mathbb{R}^{n \times q}$, $\nabla \mathbf{F}(.) \in \mathbb{R}^{d \times q}$ are the gradients of $\mathcal{L}^{t}$ and $\mathbf{F}$, and the time-dependent loss function $\mathcal{L}^{t}$ represents potential modifications of the gradient descent, for instance mini-batch sampling with batch-size being a finite fraction of $d$ in the high-dimensional limit. \newline
    
    Our main result is an asymptotically (i.e. in the high-dimensional limit) exact characterization of the distribution of the iterates $\ww^{t}$ and preactivations $\XX\ww^{t}$ at each time step. In particular, our results encompass the following special cases:
    \newline
    \begin{enumerate}
        \item an exact asymptotic characterization of discrete-time (multi-pass) stochastic gradient descent with mini-batch sizes proportional to the data dimension;
        \item first-order gradient based methods solving problem \eqref{eq:ERM_definition} with a data matrix $\XX$ with any positive definite covariance $\bm{\Sigma}\in \mathbb{R}^{d \times d}$ with bounded spectral norm;
        \item a finite number $q$ of hidden units or learners;
        \item time dependent update functions which may include stochastic effects such as mini-batch sampling, learning rate schedules and thermal noise (i.e., {\it Langevin equation}), and any differentiable regularization;
        \item momentum methods such as Polyak's heavy ball and Nesterov accelerated gradient.
    \end{enumerate}
    \section{Related works}
     Rigorous proofs of dynamical mean-field theory equations first appeared in the context of spin glasses in the works \cite{arous2001aging,ben2006cugliandolo}, who applied  large deviation theory to the paths generated by the Langevin dynamics corresponding to the Hamiltonians of the Sherrington-Kirkpatrick and spherical p-spin models. 
    
    More recently, \cite{celentano2021high} proposed a different proof for the DMFT of the high-dimensional asymptotics of first order flows for the empirical risk minimization problem \eqref{eq:ERM_definition}. This new approach was based on an approximate message passing (AMP) iteration with memory, building upon an implicit mapping between the AMP iterates and the discretized gradient flow, and using the high-dimensional concentration properties of AMP iterations, the state evolution (SE) equations.
       Our proof instead is based on iterative Gaussian conditioning, and as a consequence is simpler and more direct. Iterative Gaussian conditioning is a technique introduced in the study of SE equations for AMP iterations \cite{bayati2011dynamics, javanmard2013state, bolthausen2014iterative, berthier2020state, gerbelot2021graph}. In AMP iterations, the so-called Onsager correction applied at each time step drastically simplifies the high-dimensional effective dynamics, leading to a Markovian Gaussian process. Since gradient descent has no Onsager correction, one key aspect of the proof is to show how the dynamics may be decomposed and reformulated into asymptotically tractable memory terms and additive Gaussian processes. As a result, our proof is completely explicit  and we provide intuition on how the different terms appear in subsections \ref{subsec:grad_split}  before moving to the general case in subsection \ref{sec:gen_proof}.
       
       Our proof technique based on the iterative conditioning has important benefits as it  becomes straightforward to account for additional stochastic effects that are independent on the design matrix, notably mini-batch sampling or thermal noise, as well as potential momentum terms.
       Additionally, we  allow non-separable, time-dependent update functions, which enables to handle design matrices with arbitrary well-conditioned covariance and bounded spectral norm. 
         We do not study the continuous time limit, provided in \cite{celentano2021high} for gradient flow on separable cost functions. Notably, they prove the existence and uniqueness of the solution to the stochastic integro-differential system describing the high-dimensional gradient flow dynamics under suitable conditions. They also benefit from the universality results for AMP iterations, 
       \cite{bayati2015universality, chen2021universality}, allowing design matrices with independent sub-Gaussian entries and identity covariance. We note that the recent work \cite{montanari2022statistically} shows that non-separable first order algorithms can be reformulated as non-separable AMP iterations, building on the results of . This suggests that the proof of \cite{celentano2021high} could be adapted to the non-separable case. However, the proof of \cite{montanari2022statistically} uses an implicit mapping defined inductively, which, to the best of our knowledge, does not appear straightforward to combine with the implicit mapping from \cite{celentano2021high}.
       
       Finally, it is interesting to note that, although methods from theoretical physics are often not rigorous, a direct parallel can be drawn between our proof and derivation of the dynamical cavity method as formulated in \cite{liu2021dynamics}, \cite{mezard1987spin} and references therein for earlier appearances. Indeed, the dynamical cavity method relies on an orthogonal decomposition of the samples and iterates along a chosen direction, resulting in approximately independent Gaussian terms with different scalings. As a low dimensional projection, the term aligned with the chosen direction is of finite order, while the orthogonal component contains a number of directions proportional to the dimension and thus remains of extensive order. A Taylor expansion then allows to simplify the dynamics and obtain the DMFT equations with some algebra. In the present rigorous proof, we also perform orthogonal decompositions, but in the direction of previous iterates. For a finite number of iterations and width $q$ of the iterates, the component resulting from this projection is also of low-order, while the orthogonal component remains extensive. The proof, done by induction, then boils down to a precise control of the correlations of the different terms and concentration of various inner products appearing due to the projections using the induction hypothesis.
\section{Main result}
\label{sec:main_result}
Our main result characterizes the high-dimensional dynamics of a family of iterations that includes gradient descent iteration Eq.~\eqref{eq:grad_desc}, and takes the generic form 
    \begin{align}
        \label{eq:the_dynamics1}
        \mathbf{v}^{t+1} &= \mathbf{h}^{t}\left(\left\{\mathbf{v}^{k}\right\}_{k=0}^{t}\right)+\mathbf{X}^{\top}\mathbf{g}^{t}(\mathbf{r}^{t})  \\
        \mathbf{r}^{t} &= \mathbf{X}\sum_{k=0}^{t}\mathbf{v}^{k}\, ,
        \label{eq:the_dynamics2}
    \end{align}
initialized with $\mathbf{v}_{0} \in \mathbb{R}^{d \times q}$.
The update functions $\mathbf{g}^{t} : \mathbb{R}^{n \times q} \to \mathbb{R}^{n \times q}$ and $\mathbf{h}^{t} : \mathbb{R}^{d \times q(t+1)}\to \mathbb{R}^{d \times q}$ will belong to the regularity class of pseudo-Lipschitz functions, which will also be used to characterize the (weak) convergence of random matrices (of finite width) in the rest of the paper. This family of functions is commonly used in the AMP literature, see e.g. \cite{berthier2020state}, and its definition is reminded at definition \ref{def:pseudo-lip}.  Note that, when considering a planted model as in Eq.~(\ref{eq:ERM_definition}), the corresponding gradient based dynamics will involve a sequence of functions $\bm{g}^{t}$ implicitly depending on the data matrix $\mathbf{X}$ through the observed labels $\mathbf{y}$.
Following \cite{celentano2021high}, this additional dependence can be dealt with by considering an augmented variable $\left[\mathbf{w} \vert \mathbf{w}_{*}\right]$ and a corresponding update function involving the gradient step on $\mathbf{w}_{0}$, which is made possible by the validity of the result for matrix-valued variables of finite width. It can also be dealt with using an orthogonal decomposition in the direction of $\mathbf{w}_{*}$, see e.g. \cite{gerbelot2021graph}. We will use the former formulation to avoid redundant derivations. 
    
\subsection{Examples of algorithms belonging to the considered family}
\label{sec:algs}
    
  \paragraph{Stochastic gradient-descent} 
  
    Consider the following stochastic gradient-descent dynamics with constant step-size $\gamma$
    \begin{equation}
    \label{eq:sgd_iteration}
        \mathbf{w}^{t+1} = \mathbf{w}^{t} -\gamma\left(\frac 1b\XX^{\top}\mathbf{s}^{t} \odot \nabla \mathcal{L}(\XX\ww^{t})+\nabla \reg(\ww^{t})\right).
    \end{equation}
    where $\mathbf{s}^{t} \in \mathbb{R}^{n}$ is a random vector with i.i.d.~elements sampled at each time step according to a Bernoulli distribution with parameter $b$, and $\odot$ is the Hadamard product. This way of modelling SGD mini-batch sampling has been introduced in \cite{mignacco2020dynamical}. Now define the increment variable $\mathbf{v}^{t} = \mathbf{w}^{t}-\mathbf{w}^{t-1}$ such that, for any $t \in \mathbb{N}$, $\mathbf{w}^{t} =\sum_{k=0}^{t}\mathbf{v}^{t}$ with the convention $\mathbf{v}^{t=-1} = 0$; the preactivation term $\rr^{t} = \XX\ww^{t} \in \mathbb{R}^{n \times q}$, such that the stochastic gradient-descent iteration may be rewritten
    \begin{align}
    \label{eq:inter_grad1}
        \vv^{t+1} &= -\gamma\nabla \reg\left(\sum_{k=0}^{t}\mathbf{v}^{t}\right)-\gamma\mathbf{X}^{\top}\mathbf{s}^{t}\odot \nabla \mathcal{L}(\mathbf{r}^{t}) \\
        \mathbf{r}^{t} &= \mathbf{X}\sum_{k=0}^{t}\mathbf{v}^{t} \label{eq:inter_grad2}
    \end{align}
    which fits the form of Eq.~(\ref{eq:the_dynamics1}-\ref{eq:the_dynamics2}) by choosing $\mathbf{g}^{t}(\mathbf{r}^{t}) = -\gamma\mathbf{s}^{t} \odot \nabla \mathcal{L}(\mathbf{r}^{t})$, $\mathbf{h}^{t}(\mathbf{w}^{t}) = -\gamma\nabla \reg(\mathbf{w}^{t})$. Notice that our characterization requires that the size of the training mini-batch be a finite fraction of the full dataset.
    \paragraph{Stochastic gradient descent on cost functions involving a generic covariance}
    Consider the optimization problem 
    \begin{align}
        &\hat{{\ww}} \in \inf_{\ww \in \mathbb{R}^{d \times q}} \mathcal{L}(\XX \boldsymbol{\Sigma}^{1/2}\ww,\mathbf{y})+\reg(\ww) \\
        &\mbox{where} \thickspace \yy = \Phi_{0}\left(\XX\boldsymbol{\Sigma}^{1/2}\wwt\right),
    \end{align}
    where $\boldsymbol{\Sigma} \in \mathbb{R}^{d \times d}$ is a symmetric positive definite covariance matrix. This optimization problem can be equivalently rewritten 
    \begin{align}
        &\hat{{\ww}} \in \inf_{\ww \in \mathbb{R}^{d \times q}} \mathcal{L}(\XX \tilde{\ww},\mathbf{y})+\reg(\boldsymbol{\Sigma}^{-1/2}\tilde{\ww}) \\
        &\mbox{where} \thickspace \yy = \Phi_{0}\left(\XX\tilde{\ww}^{*}\right),
    \end{align}
    where $\tilde{\ww} = \boldsymbol{\Sigma}^{1/2} \ww, \tilde{\ww}^{*} = \boldsymbol{\Sigma}^{1/2} \ww^{*}$. Stochastic gradient descent then takes the form 
    \begin{equation}
    \label{eq:sgd_iteration_cov}
        \tilde{\mathbf{w}}^{t+1} = \tilde{\mathbf{w}}^{t} -\gamma\left(\frac 1b\XX^{\top}\mathbf{s}^{t} \odot \nabla \mathcal{L}(\XX\tilde{\ww}^{t})+\boldsymbol{\Sigma}^{-1/2}\nabla \reg(\boldsymbol{\Sigma}^{-1/2}\tilde{\ww}^{t})\right).
    \end{equation}
    The update function associated to the regularization is non-separable due to the covariance.
    \paragraph{Langevin algorithm}
    The discretized Langevin algorithm amounts to adding independent Gaussian noise to the gradient descent, leading to the following iteration
    \begin{equation}
    \label{eq:langevin_iteration}
        \mathbf{w}^{t+1} = \mathbf{w}^{t} -\gamma\left(\XX^{\top}\nabla \mathcal{L}(\XX\ww^{t})+\nabla \reg(\ww^{t})\right)+\gamma\sqrt{T}\mathbf{z}^{t}
    \end{equation}
    where $\mathbf{z}^{t} \in \mathbb{R}^{d}$ has i.i.d. standard normal elements and is independent from all other problem parameters and $\mathbf{z}^{t'}$ for all $t'\neq t$. It is then straightforward to redefine the function $\mathbf{h}^{t}(\mathbf{w}^{t}) = -\gamma\nabla \mathbf{F}(\mathbf{w}^{t})+\sqrt{T}\mathbf{z}^{t}$, which will simply lead to an additive noise with variance $T$ at each time step in the Gaussian process $u^{t}$ of the field $\nu^{t+1}$ in Corollary \ref{th:main-dmft-separable}. This modification is also observed when discretizing the DMFT equations obtained from physics methods \cite{mignacco2020dynamical}.
       
    \paragraph{Polyak momentum}
Polyak momentum \cite{polyak1964some} (or heavy-ball method) reads 
\begin{equation}
    \label{eq:polyak_iteration}
        \mathbf{w}^{t+1} = \mathbf{w}^{t} -\gamma\left(\XX^{\top}\nabla \mathcal{L}(\XX\ww^{t})+\nabla \reg(\ww^{t})\right)+\beta\left(\mathbf{w}^{t}-\mathbf{w}^{t-1}\right)
\end{equation}
with gradient step size $\alpha$ and momentum parameter $\beta$.
Using the same intermediate variables as those introduced for the reformulation of the stochastic gradient-descent iteration Eq.~\eqref{eq:sgd_iteration} into dynamics of the form of Eq.~(\ref{eq:the_dynamics1}-\ref{eq:the_dynamics2}), we obtain 
    \begin{align}
    \label{eq:inter_polyak}
        \vv^{t+1} &= -\gamma\nabla \reg(\sum_{k=0}^{t}\mathbf{v}^{t})-\gamma\mathbf{X}^{\top}\nabla \mathcal{L}(\mathbf{r}^{t})+\beta\mathbf{v}^{t} \\
        \mathbf{r}^{t} &= \mathbf{X}\sum_{k=0}^{t}\mathbf{v}^{t} \label{eq:inter_polyak2}
    \end{align}
which fits the form of Eq.~(\ref{eq:the_dynamics1}-\ref{eq:the_dynamics2}) by choosing $\mathbf{g}^{t}(\mathbf{r}^{t}) = -\gamma\nabla \mathcal{L}(\mathbf{r}^{t})$, and \\ $\mathbf{h}^{t}(\left\{\mathbf{v}^{k}\right\}_{k=0}^{t}) = -\gamma\nabla \reg(\sum_{k=0}^{t}\mathbf{v}^{k})+\beta\mathbf{v}^{t}$. \\
\par 
\paragraph{Nesterov accelerated gradient}
Nesterov accelerated gradient \cite{nesterov1983method} is defined as an iteration of three sequences parametrized by stepsizes $\tau^{t},\gamma^{t},\nu^{t},\alpha^{t}$ and initialized with $\mathbf{w}^{0},\mathbf{z}^{0}$, taking the form
\begin{align}
\label{eq:nesterov1}
    \mathbf{y}^{t} &= \mathbf{w}^{t}+\tau^{t}(\mathbf{z}^{t}-\mathbf{w}^{t}) \\
    \mathbf{w}^{t+1} &=\mathbf{y}^{t}-\gamma^{t}\left(\XX^{\top}\nabla \mathcal{L}(\XX\yy^{t})+\nabla \reg(\yy^{t})\right) \\
    \mathbf{z}^{t+1} &= \mathbf{z}^{t}+\mu^{t}\left(\mathbf{y}^{t}-\mathbf{z}^{t}\right)-\alpha^{t}\left(\XX^{\top}\nabla \mathcal{L}(\XX\yy^{t})+\nabla \reg(\yy^{t})\right)
\end{align}
Defining the variables $ \mathbf{u}^{t+1} = \mathbf{w}^{t+1}-\mathbf{w}^{t} \in \mathbb{R}^{d}, \tilde{\mathbf{u}}^{t+1} = \mathbf{z}^{t+1}-\mathbf{z}^{t} \in \mathbb{R}^{d},\mathbf{v}^{t} = \left[\mathbf{u}^{t} \vert \tilde{\mathbf{u}}^{t}\right] \in \mathbb{R}^{d \times 2},\mathbf{x}^{t} = \left[\mathbf{w}^{t} \vert \mathbf{z}^{t}\right] = \sum_{k=0}^{t}\mathbf{v}^{k} \in \mathbb{R}^{d \times 2}$, $\mathbf{r}^{t} = \mathbf{X}\sum_{k=0}^{t}\mathbf{v}^{k}$, we may fit these equations to the form of Eq.~(\ref{eq:the_dynamics1}-\ref{eq:the_dynamics2}) by defining 
\begin{align}
    &\mathbf{h}^{t}:\mathbb{R}^{d \times 2(t+1)} \to \mathbb{R}^{d\times 2} \\
    &\left\{\mathbf{v}^{k}\right\}_{k=0}^{t} \to \left[\sum_{k=0}^{t}\mathbf{v}^{k}\begin{bmatrix}-\tau^{t} \\
    \tau^{t}\end{bmatrix} \vert \sum_{k=0}^{t}\mathbf{v}^{k}\begin{bmatrix}\mu^{t}(1-\tau^{t}) \\ \mu^{t}(\tau^{t}-1) \end{bmatrix}\right]  \\
    &\hspace{1cm}+\left[-\gamma^{t}\nabla F\left(\sum_{k=0}^{t}\mathbf{v}^{k}\begin{bmatrix}1-\tau^{t} \\
    \tau^{t}\end{bmatrix}\right) \vert -\alpha^{t}\nabla F\left(\sum_{k=0}^{t}\mathbf{v}^{k}\begin{bmatrix}1-\tau^{t} \\
    \tau^{t}\end{bmatrix}\right) \right]  \\
    &\mathbf{g}^{t} : \mathbb{R}^{n \times 2} \to \mathbb{R}^{n \times 2} \\
    &\mathbf{r}^{t} \to \left[-\gamma^{t}\nabla \mathcal{L}\left(\mathbf{r}^{t}\begin{bmatrix}1-\tau^{t} \\
    \tau^{t}\end{bmatrix}\right) \vert -\alpha^{t}\nabla \mathcal{L}\left(\mathbf{r}^{t}\begin{bmatrix}1-\tau^{t} \\
    \tau^{t}\end{bmatrix}\right)\right]
\end{align}
The details of this mapping are given in Appendix \ref{sec:Nesterov_detail}.

\subsection{Statement of the main theorem}  
\quad \\
\paragraph{Notations} 

We adopt the same notations as in \cite{berthier2020state,gerbelot2021graph}.
 For two sequences of random variables $X_{n},Y_{n}$, we write $X_{n} \approxP Y_{n}$ when their difference converges in probability to $0$, i.e., $X_{n}-Y_{n} \xrightarrow[]{P}0$.
 Let $\cS_q^+$ denote the space of positive semi-definite matrices of size $q \times q$.
For any matrix $\boldsymbol{\kappa}\in \cS_{q}^{+}$ and a random matrix $\bZ\in \mathbb{R}^{N \times q}$ we write $\bZ \sim \mathbf{N}(0,\boldsymbol{\kappa} \otimes \mathbf{I}_{N})$ if $\bZ$ is a matrix with jointly Gaussian entries such that for any $1\leqslant i,j\leqslant q$, $\mathbb{E}[\bZ^{i}(\bZ^{j})^{\top}] = \boldsymbol{\kappa}_{i,j}\mathbf{I}_{N}$, where $\bZ^{i},\bZ^{j}$ denote the i-th and j-th columns of $\bZ$. The i-th line of the matrix $\bZ$ is denoted $\bZ_{i}$. If $f: \mathbb{R}^{N \times q} \to \mathbb{R}^{N \times q}$ is a function and $i \in \{1, \dots N\}$, we write $f_{i}:\mathbb{R}^{N \times q} \to \mathbb{R}^{q}$ to denote the component of $f$ generating the $i$-th line of its image, i.e., if $\mathbf{X} \in \mathbb{R}^{N \times q}$, 
\begin{equation*}
    f(\mathbf{X}) = \begin{bmatrix}
    f_{1}(\mathbf{X}) \\
    \vdots \\
    f_{N}(\mathbf{X})\end{bmatrix} \in \mathbb{R}^{N \times q} \, .
\end{equation*}
We write $\frac{\partial f_{i}}{\partial \mathbf{X}_{i}}$ the $q\times q$ Jacobian containing the derivatives of $f_{i}$ with respect to (w.r.t.) the $i$-th line $\mathbf{X}_{i}\in \mathbb{R}^{q}$:
\begin{equation}
\label{eq:Ons_jacob}
    \frac{\partial f_{i}}{\partial \mathbf{X}_{i}} = \begin{bmatrix}\frac{\partial (f_{i}(\mathbf{X}))_{1}}{\partial \mathbf{X}_{i1}} & \dots & \frac{\partial (f_{i}(\mathbf{X}))_{1}}{\partial \mathbf{X}_{iq}} \\
    \vdots& &\vdots \\
    \frac{\partial (f_{i}(\mathbf{X}))_{q}}{\partial \mathbf{X}_{i1}} & \dots & \frac{\partial (f_{i}(\mathbf{X}))_{q}}{\partial \mathbf{X}_{iq}}
    \end{bmatrix} \in \mathbb{R}^{q \times q} \, .
\end{equation} \quad \\
\\
We will also use the following class of functions to state our assumptions and convergence results.
\begin{definition}[pseudo-Lipschitz function]
    \label{def:pseudo-lip}
    For $k \in \mathbb{N}^{*}$ and any $n,m \in \mathbb{N}^{*}$, a function $\Phi : \mathbb{R}^{n \times q} \to \mathbb{R}^{m \times q}$ is said to be \emph{pseudo-Lipschitz of order k} if there exists a constant L such that for any $\mathbf{x},\mathbf{y} \in \mathbb{R}^{n \times q}$, 
    \begin{equation}
        \frac{\norm{\Phi(\mathbf{x})-\Phi(\mathbf{y})}_{F}}{\sqrt{m}} \leqslant L \left(1+\left(\frac{\norm{\mathbf{x}}_{F}}{\sqrt{n}}\right)^{k-1}+\left(\frac{\norm{\mathbf{y}}_{F}}{\sqrt{n}}\right)^{k-1}\right)\frac{\norm{\mathbf{x}-\mathbf{y}}_{F}}{\sqrt{n}}
    \end{equation}
    A family of pseudo-Lipschitz functions $\{\phi_{n}\}_{n \in \mathbb{N}}$ is said to be \emph{uniformly} pseudo-Lipschitz if the pseudo-Lipschitz constants $L_{n}$ verify $L_{n} < \infty$ for each $n$ and if $\limsup_{n \to \infty} L_{n} < \infty$. 
    \end{definition}
    We now state the required assumptions for our main result to hold. These assumptions are similar to the ones required for the proof of state evolution equations related to approximate message passing iterations with non-separable update functions and matrix valued iterates, see e.g. \cite{gerbelot2021graph}. \quad \\
    \paragraph{Assumptions}
    \begin{enumerate}[font={\bfseries},label={(A\arabic*)}]
    \item \label{main_assum_1} the dimensions of the problem $n,d$ go to infinity with finite ratio $n/d = \alpha$, where $\alpha \in (0,\infty)$;
    \item \label{main_assum_2} the matrix $\mathbf{X}$ has i.i.d. $\mathbf{N}(0,\frac{1}{d})$ elements. As we have seen at Eq.~\eqref{eq:sgd_iteration_cov}, a positive definite covariance $\boldsymbol{\Sigma}$ with bounded spectral norm can be added to the optimization problem Eq.~\eqref{eq:ERM_definition}, leading to non-separable functions in the corresponding gradient descent iteration. Non-separable functions are included in our next assumption;
    \item \label{main_assum_3} 
    \begin{enumerate}[font={\bfseries},label={(\alph*)}]
        \item \label{main_assum_3a} for any $t \in \mathbb{N}$, the functions $\mathbf{g}^{t} : \mathbb{R}^{n \times q} \to \mathbb{R}^{n \times q}, \mathbf{h}^{t}: \mathbb{R}^{d \times q} \to \mathbb{R}^{d \times q}$ are deterministic, pseudo-Lipschitz continuous of order $k$;
        \item \label{main_assum_3b} for any $t \in \mathbb{N}$, the functions $\mathbf{g}^{t} : \mathbb{R}^{n \times q} \to \mathbb{R}^{n \times q}, \mathbf{h}^{t}: \mathbb{R}^{d \times q} \to \mathbb{R}^{d \times q}$ are separable and, for any $1 \leq i \leq N$, the functions 
        $g_{i}^{t} : \mathbb{R}^{q} \to \mathbb{R}, h_{i}^{t}:\mathbb{R}^{q} \to \mathbb{R}$ are defined by
        \begin{align*}
            g_{i}^{t} = \tilde{g}(x,b_{i}^{t}) \\
            h_{i}^{t} = \tilde{h}(x,d_{i}^{t})
        \end{align*}
        where $\tilde{g} : \mathbb{R}^{q+1} \to \mathbb{R},\tilde{h}:\mathbb{R}^{q+1} \to \mathbb{R}$ are pseudo-Lipschitz of order $k$, and, for any $t \in \mathbb{N}$, $\mathbf{b}^{t}, \mathbf{d}^{t} \in \mathbb{R}^{N}$ are 
        random vectors with i.i.d. sub-gaussian entries, independent from the random matrix $\mathbf{X}$ and from $\mathbf{b}^{t'},\mathbf{d}^{t'}$ for any $t' \neq t$.
    \end{enumerate}
    \item \label{main_assum_4} the initialization $\mathbf{v}_{0}$ is deterministic and $\frac{1}{d}\langle \mathbf{v}_{0}, \mathbf{v}_{0} \rangle$ converges to a finite constant as $d \to \infty$;
    \item \label{main_assum_5} the following limit exists and is finite:
    \begin{equation*}
        \lim_{d \to \infty} \frac{1}{d} \langle \mathbf{h}^{0}(\mathbf{v}^{0}), \mathbf{h}^{0}(\mathbf{v}^{0}) \rangle 
    \end{equation*}
    \item \label{main_assum_6} for any $t>0$, let $\{\kappa_{kl}\}_{0 \leqslant k,l \leqslant t}$ be an array of deterministic $q \times q$ positive definite matrices with bounded spectral norm and let $\mathbf{Z}^{0},\mathbf{Z}^{1},...,\mathbf{Z}^{t}$ be a sequence of $d \times q$ random matrices such that $(\mathbf{Z}^{0},\mathbf{Z}^{1},...,\mathbf{Z}^{t}) \sim \mathbf{N}(0,\{\kappa_{kl}\}_{0 \leqslant k,l \leqslant t} \otimes \mathbf{I}_{d})$. The following limit exists and is finite:
    \begin{equation*}
        \lim_{d \to \infty} \frac{1}{d} \mathbb{E}\left[\langle \mathbf{h}^{0}(\mathbf{v}^{0}),\mathbf{h}^{t}\left(\left\{\mathbf{Z}^{k}\right\}_{k=0}^{t}\right) \rangle\right]
    \end{equation*} \label{assum:cov1}
    \item \label{main_assum_7} for any $t >0$, define the sequence of random matrices $\mathbf{Z}^{0},\mathbf{Z}^{1},...,\mathbf{Z}^{t}$ as in \ref{assum:cov1}. For any $s,t>0$, let $\tilde{\boldsymbol{\kappa}}_{st}$ be a deterministic, $2q \times 2q$
    positive definite matrix with bounded spectral norm and $\tilde{\mathbf{Z}}^{s},\tilde{\mathbf{Z}}^{t}$ two $n \times q$ random matrices such that $(\tilde{\mathbf{Z}}^{s},\tilde{\mathbf{Z}}^{t}) \sim \mathbf{N}(0,\tilde{\boldsymbol{\kappa}}_{st} \otimes \mathbf{I}_{n})$. The following limits exist and are finite:
    \begin{align*}
        &\lim_{d \to \infty} \frac{1}{d}\mathbb{E}\left[\langle \mathbf{h}^{s}\left(\left\{\mathbf{Z}^{k}\right\}_{k=0}^{s}\right), \mathbf{h}^{t}\left(\left\{\mathbf{Z}^{k}\right\}_{k=0}^{t}\right) \rangle\right] \\
        &\lim_{n \to \infty} \frac{1}{n}\mathbb{E}\left[\langle \mathbf{g}^{s}\left(\tilde{\mathbf{Z}}^{s}\right), \mathbf{g}^{t}\left(\tilde{\mathbf{Z}}^{t}\right) \rangle\right]
    \end{align*}
    \end{enumerate}
     Our main result is presented in the following theorem: 
    \begin{theorem}(High-dimensional dynamics of gradient-based methods)
        \label{th:main_dmft} 
        Consider the following discrete time stochastic process
        \begin{align}
            \label{eq:main_th_eq1}
            \bm{\nu}^{t+1}&= \bm{\theta}^{t}{\Gamma}^{t}+\mathbf{h}^{t}\left(\left\{\bm{\nu}^{k}\right\}_{k=0}^{t}\right)+\sum_{k=0}^{t-1}\bm{\theta}^{k}{R}_{g}(t,k)+\bm{u}^{t} \in \mathbb{R}^{d \times q} \\ \bm{\theta}^{t} &= \sum_{k=0}^{t}\bm{\nu}^{k} \in \mathbb{R}^{d \times q} \\
  \bm{\eta}^{t} &= \sum_{k=0}^{t-1}\bm{g}^{k}(\bm{\eta}^{k}){R}_{\theta}(t,k)+\bm{\omega}^{t} \in \mathbb{R}^{n \times q} \\
    {R}_{\theta}(t,s) &= \lim_{d \to \infty} \frac{1}{d} \sum_{i=1}^{d}\mathbb{E}\left[\frac{\partial \theta^{t}_{i}}{\partial u^{s}_{i}}\right] \in \mathbb{R}^{q \times q} \\
      R_{g}(t,s) &= \lim_{d \to \infty} \frac{1}{d}\sum_{i=1}^{n}\mathbb{E}\left[\frac{\partial {\bar{g}}^{t}_{i}}{\partial \omega_{i}^{s}}(\bm{\eta}^{t})\right] \in \mathbb{R}^{q \times q} \\
            \Gamma^{t} &= \lim_{d \to \infty} \frac{1}{d}\sum_{i=1}^{n}\mathbb{E}\left[\frac{\partial {g}^{t}_{i}}{\partial \eta^{t}_{i}}(\bm{\eta}^{t})\right] \in \mathbb{R}^{q \times q} \\
           C_{\theta}(t,s) &= \lim_{d \to \infty} \frac{1}{d}\mathbb{E}\left[\left(\bm{\theta}^{t}\right)^{\top}\bm{\theta}^{s}\right] \in \mathbb{R}^{q \times q} \\
            C_{g}(t,s) &= \lim_{d \to \infty} \frac{1}{d}\mathbb{E}\left[\mathbf{g}^{s}(\bm{\eta}^{s})^{\top}\mathbf{g}^{t}(\bm{\eta}^{t})\right] \in \mathbb{R}^{q \times q} \label{eq:main_th_eq2}
        \end{align}
        initialized with $\bm{\nu}^{0} = \mathbf{v}^{0}$, where $\bm{u}^{t},\bm{\omega}^{t}$ have i.i.d. lines in $\mathbb{R}^{q}$ which are Gaussian processes with covariances $C_{g}^{s,t},  C_{\theta}^{s,t}$. In the above, the notation $\frac{\partial {\bar{g}}^{t}_{i}}{\partial \omega_{i}^{s}}(\bm{\eta}^{t})$ denotes the partial derivative of $\bar{\mathbf{g}}^{t}(\bomega^{1:t-1}) = \mathbf{g}^{t}(\beeta^{t})$ considered as a function of the $\{\bomega^{k}\}_{1\leqslant k \leqslant t-1}$. Consider the iteration Eq.~(\ref{eq:the_dynamics1}-\ref{eq:the_dynamics2}). Then, under assumptions \ref{main_assum_1}-\ref{main_assum_3}.\ref{main_assum_3a},\ref{main_assum_4}-\ref{main_assum_7}; or under assumptions \ref{main_assum_1}-\ref{main_assum_3}.\ref{main_assum_3b},\ref{main_assum_4}, for any $t \in \mathbb{N}$, and any pseudo-Lipschitz functions $\Psi : \mathbb{R}^{d \times q(t+1)} \to \mathbb{R}$ and $\Phi:\mathbb{R}^{n \times qt} \to \mathbb{R}$:
        \begin{equation}
           \begin{split} &\Psi(\mathbf{w}^{0}, ..., \mathbf{w}^{t}) \approxP \mathbb{E}\left[\Psi(\bm{\theta}^{0}, ...,\bm{\theta}^{t})\right]; \thickspace \mbox{and}\\ & \Phi(\mathbf{r}^{0}, ..., \mathbf{r}^{t-1}) \approxP \mathbb{E}\left[\Phi(\bm{\eta}^{0}, ..., \bm{\eta}^{t-1})\right]  .  
     \end{split}\end{equation}
    \end{theorem}
    Note that, even if the effective dynamics are written as a high-dimensional recursion in the non-separable case, all the Gaussian fields have i.i.d. variables (in $\mathbb{R}^{q}$) that are independent across their extensive dimensions, and are completely parametrized by low-dimensional quantities. The limits defining the discrete time process in Theorem \ref{th:main_dmft} are all well-defined under Assumptions \ref{main_assum_1}-\ref{main_assum_7} (choosing either (A3.a) or (A3.b)),  and are explicitly controlled in the induction step of the proof. \quad \\
     \par
    In the separable case, i.e. under assumptions \ref{main_assum_1}-\ref{main_assum_3}.\ref{main_assum_3b},\ref{main_assum_4}, Eq.\eqref{eq:main_th_eq1}-\eqref{eq:main_th_eq2} simplify and the resulting discrete time process can be written explicitly as a low-dimensional recursion. The following corollary gives the corresponding low-dimensional limiting dynamics for the SGD iteration described at Eq.~\eqref{eq:sgd_iteration}, under the following assumptions :
    \begin{enumerate}[font={\bfseries},label={(B\arabic*)}]
        \item \label{main_assum_b1} the dimensions of the problem $n,d$ go to infinity with finite ratio $n/d = \alpha$, where $\alpha \in (0,\infty)$;
        \item \label{main_assum_b2} the matrix $\mathbf{X}$ has i.i.d. $\mathbf{N}(0,\frac{1}{d})$ elements.
        \item \label{main_assum_b3} the loss function $\mathcal{L}$ and regularization $\mathbf{F}$ are differentiable and separable with respective component-wise functions $l:\mathbb{R}^{q} \to \mathbb{R},f:\mathbb{R}^{q} \to \mathbb{R}$. $\mathcal{L}$ is twice differentiable, and the 
        gradients $l':\mathbb{R}^{q} \to \mathbb{R}^{q}, f':\mathbb{R}^{q} \to \mathbb{R}^{q}$ are pseudo-Lipschitz of order $k$.
        \item \label{main_assum_b4} the initialization $\mathbf{v}_{0}$ is deterministic and $\frac{1}{d}\langle \mathbf{v}_{0}, \mathbf{v}_{0} \rangle$ converges to a finite constant as $d \to \infty$;
    \end{enumerate}
    \begin{corollary}
        \label{th:main-dmft-separable}Consider the SGD iteration of Eq.~\eqref{eq:sgd_iteration}. Let $l''$ denote the $q\times q$ Hessian of $l$.
        Consider the following discrete-time stochastic process 
        \begin{align}
        \label{eq:effective_nu}
            \nu^{t+1}&= \Gamma^{t}\theta^{t}-\gamma f'(\theta^{t})+\sum_{k=0}^{t-1}R_{g}(t,k)\theta^{k}+u^{t} \in \mathbb{R}^{q} \\
            \theta^{t} &= \sum_{k=0}^{t}\nu^{k} \in \mathbb{R}^{q} \\
            \eta^{t} &= -\gamma\sum_{k=0}^{t-1}R_{\theta}(t,k)s^{k}l'(\eta^{k})+\omega^{t} \in \mathbb{R}^{q} \\
            R_{\theta}(t,s) &=  \mathbb{E}\left[\frac{\partial \theta^{t}}{\partial u^{s}}\right] \in \mathbb{R}^{q \times q} \\
            R_{g}(t,s) &= -\alpha\gamma \mathbb{E}\left[s^{t}\frac{\partial \bar{l^{'}}}{\partial \omega^{s}}(\eta^{t})\right] \in \mathbb{R}^{q \times q} \\
            \Gamma^{t} &= -\alpha\gamma \mathbb{E}\left[s^{t}l^{''}(\eta^{t})\right] \in \mathbb{R}^{q \times q} \\
            C_{\theta}(t,s) &= \mathbb{E}\left[\theta^{s}\left(\theta^{t}\right)^{\top}\right] \in \mathbb{R}^{q \times q} \\
            C_{g}(t,s) &= \alpha\gamma^{2} \mathbb{E}\left[s^{s}s^{t}l^{'}(\eta^{s})l^{'}(\eta^{t})^{\top}\right] \in \mathbb{R}^{q \times q}
        \end{align}
        initialized with $\bnu^{0} = \mathbf{v}^{0}$, where $u^{t},\omega^{t}$ are Gaussian processes in $\mathbb{R}^{q}$ with covariances $C_{g}({s,t})$, \\$  C_{\theta}({s,t})$. In the above, $\frac{\partial \bar{l^{'}}}{\partial \omega^{s}}(\eta^{t})$ denotes the partial derivative of $\bar{l^{'}}(\omega^{1:t-1}) = l'(\eta^{t})$ considered as a function of the $\{\omega^{k}\}_{1 \leqslant k \leqslant t-1}$.
        Then, under assumptions \ref{main_assum_b1}-\ref{main_assum_b4}, for any $t \in \mathbb{N}$, and any 
        pseudo-Lipschitz functions $\psi: \mathbb{R}^{q(t+1)} \to \mathbb{R}$ and $\phi: \mathbb{R}^{qt} \to \mathbb{R}$: 
        \begin{align}
            &\frac{1}{d}\sum_{i=1}^{d}\psi((\mathbf{w}^{0}, ..., \mathbf{w}^{t})_{i}) \xrightarrow[n,d \to 
            \infty]{P} \mathbb{E}\left[\psi(\theta^{0}, ...,\theta^{t})\right], \\ &\frac{1}{n}\sum_{j=1}^{n}\phi((\mathbf{r}^{0}, ..., \mathbf{r}^{t-1})_{i}) \xrightarrow[n,d \to \infty]{P} \mathbb{E}\left[\phi(\eta^{0}, ..., \eta^{t-1})\right]
        \end{align}
    \end{corollary}

The proof of Corollary \ref{th:main-dmft-separable} can be found in appendix \ref{app:proof_coroll}. We remind that, to obtain the correlation with a planted vector $\bm{w}^{*}$ as in problem \ref{eq:ERM_definition}, we may use the same mapping from section 4.1 of \cite{celentano2021high}. 

As another concrete example, we give the equations for the case of stochastic gradient descent on a cost function involving a non-identity covariance corresponding to Eq.~\eqref{eq:sgd_iteration_cov}, where we assume that the loss function $\mathcal{L}$ is separable and twice differentiable. The equations defining the field $\btheta^{t}$ become non-separable, leading to:
\begin{align}
            \bm{\nu}^{t+1}&= \bm{\theta}^{t}{\Gamma}^{t}+\boldsymbol{\Sigma}^{-1/2}\nabla\mathbf{F}\left(\boldsymbol{\Sigma}^{-1/2}\bm{\theta}^{t}\right)+\sum_{k=0}^{t-1}\bm{\theta}^{k}{R}_{g}(t,k)+\bm{u}^{t} \in \mathbb{R}^{d \times q} \\
            \bm{\theta}^{t} &= \sum_{k=0}^{t}\bm{\nu}^{k} \in \mathbb{R}^{d \times q} \\
  \eta^{t} &= -\gamma\sum_{k=0}^{t-1}R_{\theta}(t,k)s^{k}l'(\eta^{k})+\omega^{t} \in \mathbb{R}^{q} \\
    {R}_{\theta}(t,s) &= \lim_{d \to \infty} \frac{1}{d} \sum_{i=1}^{d}\mathbb{E}\left[\frac{\partial \theta^{t}_{i}}{\partial u^{s}_{i}}\right] \in \mathbb{R}^{q \times q} \\
            R_{g}(t,s) &= -\alpha\gamma \mathbb{E}\left[s^{t}\frac{\partial l^{'}}{\partial \omega^{s}}(\eta^{t})\right] \in \mathbb{R}^{q \times q} \\
            \Gamma^{t} &= -\alpha\gamma \mathbb{E}\left[s^{t}l^{''}(\eta^{t})\right] \in \mathbb{R}^{q \times q} \\
            C_{g}(t,s) &= \alpha\gamma^{2} \mathbb{E}\left[s^{s}s^{t}l^{'}(\eta^{s})l^{'}(\eta^{t})^{\top}\right] \in \mathbb{R}^{q \times q} \\
            C_{\theta}(t,s) &= \lim_{d \to \infty} \frac{1}{d}\mathbb{E}\left[\left(\bm{\theta}^{t}\right)^{\top}\bm{\theta}^{s}\right] \in \mathbb{R}^{q \times q} 
        \end{align}
Note that in this case, $\btheta^{t}$ describes the field $\tilde{\bm{w}}^{t} = \boldsymbol{\Sigma}^{1/2}\bm{w}^{t}$. To recover the properties of $\bm{w}^{t}$, we may simply apply $\boldsymbol{\Sigma}^{-1/2}$ to $\btheta^{t}$ which will conserve the pseudo-Lipschitz property of any low-dimensional such observable owing to the positive definiteness assumption on $\boldsymbol{\Sigma}$ and its bounded spectral norm.
\section{Proof}
In the next two subsections, we provide intuition on our proof method. Subsection \ref{subsec:grad_split} gives the exact asymptotic characterization of a gradient descent iteration with no regularization and a sample splitting assumption, where a fresh data matrix is drawn at each time step. This drastically simplifies the analysis and gives a simple result that is straightforward to interpret. We note that gradient-descent with sample-splitting was recently studied in \cite{chandrasekher2021sharp} using Gaussian comparison inequalities. We then move to the generic case, proving Theorem \ref{th:main_dmft} using an induction on the variables $\bm{r}^{t},\bm{u}^{t+1}$. The full induction step for $\bm{r}^{t}$ is given in the main text, while the induction step on $\bm{u}^{t+1}$, the structure of which is similar, is deferred to Appendix \ref{sec:app_proof}. Useful intermediate lemmas are gathered in Appendix \ref{app:tech_app}. \quad \\
\paragraph{Notations for the proof}
We will use the following additional notations in the proof : 
for two random variables $X$ and $Y$, and a $\sigma$-algebra $\mathfrak{S}$, we will also use $X\vert_{\mathfrak{S}} \stackrel{d} = Y$ to mean that for any integrable function $\phi$ and any $\mathfrak{S}$-measurable bounded random variable $Z$, $\mathbb{E}\left[\phi(X)Z\right] = \mathbb{E}\left[\phi(Y)Z\right]$.  We use $\mathbf{I}_{N}$ to denote the $N \times N$ identity matrix, and $0_{N \times N}$ the $N \times N$ matrix with zero entries. We use $\sigma_{\min}(\mathbf{Q})$ and $\sigma_{\max}(\mathbf{Q}) = \norm{\mathbf{Q}}_{{\rm op}}$ to denote the minimum and maximum singular values of a given matrix $\mathbf{Q}$. For two matrices $\mathbf{Q}$ and $\mathbf{P}$ with the same number of rows, we denote their horizontal concatenation with $\left[\mathbf{P} \vert \mathbf{Q}\right]$. The orthogonal projector onto the range of a given matrix $\mathbf{M}$ is denoted $\mathbf{P}_{\mathbf{M}}$, and let $\mathbf{P}_{\mathbf{M}}^{\perp} = \mathbf{I}-\mathbf{P}_{\mathbf{M}}$. 
 
\subsection{A first example: gradient descent with sample splitting}
\label{subsec:grad_split}
Under the sample splitting assumption, the gradient descent iteration reads (for $q=1$):
\begin{equation}
    \label{eq:grad_sample_split}
    \forall t \in \mathbb{N}^{*} \quad \mathbf{w}^{t+1} = \mathbf{w}^{t}-\gamma^{t} (\mathbf{A}^{t})^{\top}\nabla \mathbf{f}(\mathbf{A}^{t}\mathbf{w}^{t}) 
\end{equation}
where, for any $t \in \mathbb{N}$, $\mathbf{A}^{t} \in \mathbb{R}^{n \times d}$ is a matrix with i.i.d. Gaussian elements and variance $1/d$ independent on all other $\left\{\mathbf{A}^{i}\right\}_{i\neq t}$, $\gamma^{t} \in \mathbb{R}$ is a scalar step-size and $\mathbf{f}$ is a twice differentiable, deterministic function
with pseudo-Lipschitz gradient $\nabla \mathbf{f} : \mathbb{R}^{n} \to \mathbb{R}^{n}$. We also assume that $\mathbf{f}$ is separable, with an elementwise operation denoted $f$. The iteration is initialized with $\mathbf{w}^{0} \in \mathbb{R}^{d}$, a random vector independent on $\mathbf{A}$ with i.i.d. subgaussian elements.
Starting at $t=0$, we condition equation \eqref{eq:grad_sample_split} on (the sigma algebra generated by) $\mathbf{w}^{0},\mathbf{A}^{0}\mathbf{w}^{0}$, and obtain, using lemma \ref{lemma:cond_lemma}:
\begin{align}
    \mathbf{w}^{1}\vert_{\mathbf{w}^{0},\mathbf{A}^{0}\mathbf{w}^{0}} & \stackrel{d}= \mathbf{w}^{0}-\gamma^{0}\left(\mathbf{A}^{0}\mathbf{P}_{\mathbf{w}^{0}}+\tilde{\mathbf{A}}^{0}\mathbf{P}^{\perp}_{\mathbf{w}^{0}}\right)^{\top}\nabla \mathbf{f}(\mathbf{A}^{0}\mathbf{w}^{0}) \\
    & = \mathbf{w}^{0}-\gamma^{0}\mathbf{w}^{0}\frac{1}{\norm{\mathbf{w}^{0}}_{2}^{2}}\left(\mathbf{A}^{0}\mathbf{w}^{0}\right)^{\top}\nabla \mathbf{f}(\mathbf{A}^{0}\mathbf{w}^{0})-\gamma^{0}\mathbf{P}^{\perp}_{\mathbf{w}^{0}}(\tilde{\mathbf{A}}^{0})^{\top}\nabla \mathbf{f}(\mathbf{A}^{0}\mathbf{w}^{0})\, .
\end{align}
where $\tilde{\mathbf{A}}^{0}$ is a copy of $\mathbf{A}^{0}$ whose elements are independent of those of $\mathbf{A}^{0}$ and of the elements of $\mathbf{w}^{0}$.
Since, by assumption, the entries of $\mathbf{w}_{0}$ are independent on the entries of $\mathbf{A}_{0}$, conditionally on $\mathbf{w}_{0}$, me may lift the conditioning on $\mathbf{A}\mathbf{w}_{0}$ to obtain that the vector $\mathbf{A}^{0}\mathbf{w}^{0}$ has i.i.d. entries distributed according to $\mathcal{N}(0,\frac{1}{d}\norm{\mathbf{w}^{0}}^{2}_{2})$. We can then write  
\begin{equation}
    \frac{1}{\norm{\mathbf{w}^{0}}_{2}^{2}}\left(\mathbf{A}^{0}\mathbf{w}^{0}\right)^{\top}\nabla \mathbf{f}(\mathbf{A}^{0}\mathbf{w}^{0}) = \frac{1}{\frac{1}{d}\norm{\mathbf{w}^{0}}_{2}^{2}}\frac{1}{d}\left(\mathbf{A}^{0}\mathbf{w}^{0}\right)^{\top}\nabla \mathbf{f}(\mathbf{A}^{0}\mathbf{w}^{0})\, .
\end{equation}
Since the entries of $\mathbf{w}_{0}$ are subgaussian, $\frac{1}{d}\norm{\mathbf{w}_{0}}_{2}^{2}$ is a sum of i.i.d. subexponential random variables and thus converges almost surely to a finite constant owing to Bernstein's inequality. We can then use lemma \ref{lemma:pseudo-lip-conv} and \ref{conv_lemmas_app}, the continuous mapping theorem, and Stein's lemma to show that there exists a 
random variable $z^{0} \sim \mathcal{N}(0,\rho^{0})$ such that
\begin{equation}
    \frac{1}{\norm{\mathbf{w}^{0}}_{2}^{2}}\left(\mathbf{A}^{0}\mathbf{w}^{0}\right)^{\top}\nabla \mathbf{f}(\mathbf{A}^{0}\mathbf{w}^{0}) \approxP \alpha\mathbb{E}\left[f''(z^{0})\right]
\end{equation}
where $\rho^{0} = \lim_{d \to \infty} \frac{1}{d}\norm{\mathbf{w}^{0}}_{2}^{2}$. Turning to the part orthogonal to $\mathbf{w}^{0}$ and using the fact that the projector $\mathbf{P}_{\mathbf{w}^{0}}$ is of rank $1$, the elements of $\tilde{\mathbf{A}}$ have variance $\frac{1}{d}$ and $\norm{\mathbf{w}^{0}}_{2}^{2} = O(d)$, lemma \ref{conv_lemmas_app} shows that 
\begin{equation}
    \frac{1}{\sqrt{d}}\norm{\mathbf{P}^{\perp}_{\mathbf{w}^{0}}\tilde{\mathbf{A}}^{\top}\nabla \mathbf{f}(\mathbf{A}^{0}\mathbf{w}^{0})-(\tilde{\mathbf{A}}^{0})^{\top}\nabla \mathbf{f}(\mathbf{A}^{0}\mathbf{w}^{0})}_{2} \approxP 0
\end{equation}
where $(\tilde{\mathbf{A}}^{0})^{\top}\nabla \mathbf{f}(\mathbf{A}^{0}\mathbf{w}^{0})$ is a vector with i.i.d elements distributed as $\mathcal{N}(0,\frac{1}{d}\norm{\nabla \mathbf{f}(\mathbf{A}^{0}\mathbf{w}^{0})}_{2}^{2})$. Once again, the function 
$\frac{1}{d}\norm{\nabla \mathbf{f}(.)}_{2}^{2}$ is scalar valued and pseudo-Lipschitz, thus lemma \ref{lemma:pseudo-lip-conv} and the continuous mapping theorem show that there exists a Gaussian random variable $u^{0} \sim \mathcal{N}(0,\tau_{0})$ such that, for any pseudo-Lipschitz function $\psi : \mathbb{R} \to \mathbb{R} $ of order $2$, 
\begin{equation}
    \frac{1}{d}\sum_{i=1}^{d}\psi(\left(\mathbf{P}^{\perp}_{\mathbf{w}^{0}}\tilde{\mathbf{A}}^{\top}\nabla \mathbf{f}(\mathbf{A}^{0}\mathbf{w}^{0})\right)_{i}) \approxP \mathbb{E}\left[\psi(u^{0})\right]
\end{equation}
where we have introduced $\tau_{0} = \lim_{n,d \to \infty} \frac{1}{d}\norm{\nabla \mathbf{f}(\mathbf{A}^{0}\mathbf{w}^{0})}_{2}^{2} = \alpha\mathbb{E}\left[(f'(z^{0}))^{2}\right]$. Using these results, 
we may now lift the conditioning and use the definition of pseudo-Lipschitz function to recover the scalar equation describing the high-dimensional behaviour of $\mathbf{w}^{1}$. A straightforward induction 
shows that, for any $t \in \mathbb{N}$, the quantity $\frac{1}{d}\norm{\mathbf{w}^{t}}_{2}^{2}$ is almost surely bounded, and the same conditioning argument can be applied along the sample splitting assumption to reach the following theorem :
\begin{theorem}(High-dimensional dynamics of gradient descent with sample splitting)
    Consider the iteration Eq.~\eqref{eq:grad_sample_split} with its set of assumptions described above. Define the following discrete-time one-dimensional stochastic process, initialized with a 
    subgaussian random variable $\omega^{0}$ with variance $\rho^{0}$:
    \begin{equation}
        \omega^{t+1} = \left(1-\gamma^{t}\alpha\mathbb{E}\left[f''(z^{t})\right]\right)\omega^{t}+\gamma^{t}u^{t}
    \end{equation}
    where $\rho^{t} = \mathbb{E}\left[(\omega^{t})^{2}\right]$, $\tau^{t} = \alpha\mathbb{E}\left[(f'(z^{t}))^{2}\right]$. $z^{t}, u^{t}$ are independent normal random variables with zero mean and respective variances $\rho^{t},\tau^{t}$.
    Then, for any $t \in \mathbb{N}$ and any pseudo-Lipschitz function of order $2$ $\psi : \mathbb{R} \to \mathbb{R}$ , the following holds 
    \begin{equation}
         \frac{1}{d}\sum_{i=1}^{d}\psi(w^{t}_{i}) \xrightarrow[n,d \to 
            \infty]{P} \mathbb{E}\left[\psi(\omega^{t})\right]
    \end{equation}
\end{theorem} 
We have obtained a full description of the asymptotic distribution of $\mathbf{w}^{t}$ in terms of a scalar equation. The sample splitting assumption however, is unrealistic. Let us move to the 
generic case that corresponds to the usual gradient descent.
\begin{figure}
    \centering
    \includegraphics[scale=0.47]{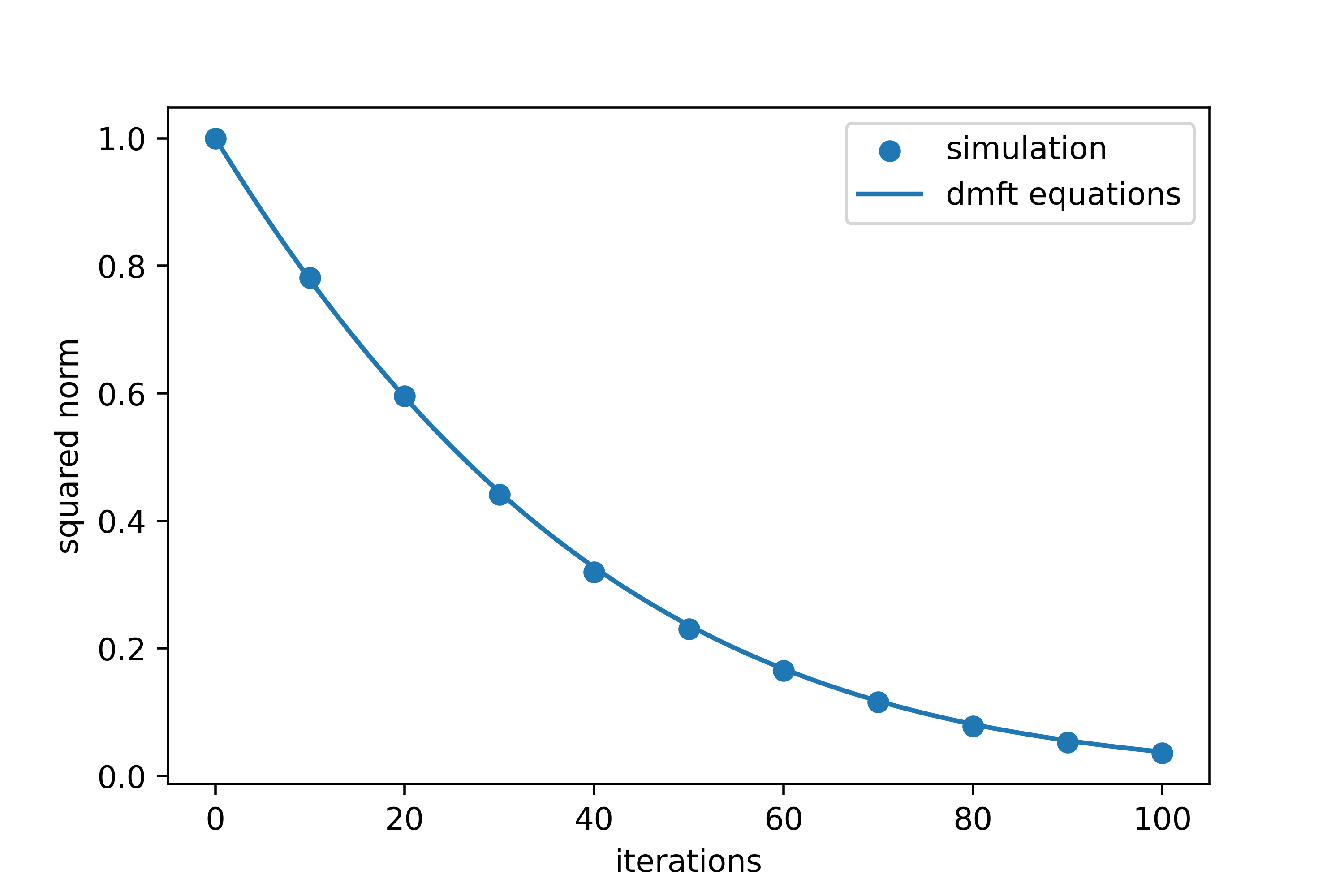}
    \caption{Gradient descent with sample splitting where $f'(z) = \mbox{tanh}(z)$ Due to the regularity of the update function and sample splitting assumption, the concentration is very fast and almost perfect matching is obtained between the theoretical and empirical curves with low dimensions (n=50,d=100) and no averaging.}
    \label{fig:mnist}
        \vspace{-0.2cm}
\end{figure}
\subsection{The general case}
\label{sec:gen_proof}
Without the sample splitting assumption, the iterates $\mathbf{x}^{t}$ and the design matrix $\mathbf{X}$ are correlated at each time step and thus there is no simple concentration towards a markovian model. We need to account for the correlation beyond the previous time step, leading to the appearance of memory kernels.
Recall the dynamics (\ref{eq:the_dynamics1}-\ref{eq:the_dynamics2}), where we introduce an additional intermediate variable $\mathbf{m}^{t} = \mathbf{g}(\mathbf{r}^{t})$:
    \begin{align}
        \label{eq:the_dynamics1_app}
        \mathbf{v}^{t+1} &= \mathbf{h}^{t}(\left\{\mathbf{v}^{k}\right\}_{k=0}^{t})+\mathbf{X}^{\top}\mathbf{m}^{t} \\
        \mathbf{m}^{t} &= \mathbf{g}^{t}(\mathbf{r}^{t}) \\
        \mathbf{r}^{t} &= \mathbf{X}\sum_{k=0}^{t}\mathbf{v}^{t}
        \label{eq:the_dynamics2_app}
    \end{align}

The proof is done by induction on $t$, and written explicitly for the set of assumptions \ref{main_assum_1}-\ref{main_assum_3}.\ref{main_assum_3a},\ref{main_assum_4}-\ref{main_assum_7}. The proof under assumptions \ref{main_assum_1}-\ref{main_assum_3}.\ref{main_assum_3b},\ref{main_assum_4} only differs
in the use of lemma \ref{lemma:sep_conc} instead of lemma \ref{lemma:pseudo-lip-conv}. 
        \paragraph{Initialization}
        At initialization, we have 
        \begin{align}
            \mathbf{v}^{0} &= \mathbf{w}^{0} \sim P_{\mathbf{v}_{0}} \quad \mbox{by definition} \quad \mathbf{v}^{0} = \bnu^{0} 
        \end{align}
        Now, using the independence of the elements of $\mathbf{X}$ on those of $\mathbf{w}_{0}$ and the fact that the elements of $\mathbf{w}_{0}$ are i.i.d. sub-gaussian, we may use lemma \ref{conv_lemmas_app} conditionally on the entries of $\mathbf{v}^{0}$ to show that there exists a Gaussian random matrix 
        $\beeta^{0} \in \mathbb{R}^{n \times q}$ with a covariance structure corresponding to $\beeta^{0} \sim \mathcal{N}(0,C_{\theta}(0,0)\otimes \mathbf{I}_{n})$ where $C_{\theta}(0,0) = \lim_{d \to \infty} \frac{1}{d}\mathbb{E}\left[(\mathbf{v}^{0})^{\top}\mathbf{v}^{0}\right]$, such that 
        \begin{align}
            \label{inter:eta0}
            \frac{1}{\sqrt{n}}\norm{\mathbf{X}\mathbf{v}^{0}-\beeta^{0}}_{F} \xrightarrow[n,d \to \infty]{P} 0
        \end{align}
        and the $q \times q$ covariance matrix $C_{\theta}(0,0)$ coincides with the one from Theorem \ref{th:main_dmft}. \quad \\
        \quad \\
        For clarity, we also include explicitly the step for $\mathbf{v}^{1}$ in the initialization of the induction, as it is the first step where a memory kernel starts to appear. By definition of the iteration, 
        \begin{align}
            \mathbf{v}^{1} = \bm{h}^{0}\left(\mathbf{v}^{0}\right)+\mathbf{X}^{\top}\mathbf{m}^{0}.
        \end{align}
        Since the functions $h^{0},g^{0}$ are continuous, $\mathbf{h}^{0}(\mathbf{v}^{0})$ and $\mathbf{m}^{0}$ are $\mathfrak{S}^{0} = \sigma\left(\mathbf{v}^{0},\mathbf{r}^{0}\right)$ measurable. Conditioning on $\mathfrak{S}^{0}$  and using lemma \ref{lemma:cond_lemma} then yields
        \begin{align}
            \mathbf{v}^{1}\vert_{\mathfrak{S}^{0}} &\stackrel{d}= \bm{h}^{0}(\mathbf{v}^{0})+\left(\mathbf{X}\vert_{\mathfrak{S}^{0}}\right)^{\top}\mathbf{m}^{0} \\
            &\stackrel{d}= \bm{h}^{0}(\mathbf{v}^{0})+\left(\mathbf{P}_{\mathbf{v}^{0}}\mathbf{X}^{\top}+\mathbf{P}^{\perp}_{\mathbf{v}^{0}}\tilde{\mathbf{X}}^{\top}\right)\mathbf{m}^{0} \\
            &= \bm{h}^{0}(\mathbf{v}^{0})+\mathbf{v}^{0}\left((\mathbf{v}^{0})^{\top}\mathbf{v}^{0}\right)^{-1}(\mathbf{v}^{0}\mathbf{X})^{\top}g^{0}(\mathbf{r}^{0})+\mathbf{P}^{\perp}_{\mathbf{v}^{0}}\tilde{\mathbf{X}}^{\top}\mathbf{m}^{0} \label{eq:decomp_0}
        \end{align}
        where $\tilde{\mathbf{X}}$ is a copy of $\mathbf{X}$ whose elements are independent on $\mathbf{X}$ and $\mathfrak{S}^{0}$.
        The middle term can be rewritten 
        \begin{equation}
            \mathbf{v}^{0}\left(\frac{1}{d}(\mathbf{v}^{0})^{\top}\mathbf{v}^{0}\right)^{-1}\frac{1}{d}(\mathbf{v}^{0}\mathbf{X})^{\top}g^{0}(\mathbf{X}\mathbf{v}^{0})
        \end{equation}
        We can then invoke lemma \ref{lemma:pseudo-lip-conv} and Eq.\eqref{inter:eta0} to obtain that 
        \begin{equation}
            \norm{\frac{1}{d}(\mathbf{v}^{0}\mathbf{X})^{\top}g^{0}(\mathbf{X}\mathbf{v}^{0})-\frac{1}{d}\mathbb{E}\left[(\beeta^{0})^{\top}\mathbf{g}^{0}(\beeta^{0})\right]}_{F} \xrightarrow[n,d \to \infty]{P} 0
        \end{equation}
        Recalling that $\beeta^{0} \sim \mathcal{N}(0,C_{\theta}(0,0)\otimes \mathbf{I}_{n})$, the matrix valued Stein's lemma \ref{matrix-stein} shows that 
        \begin{equation}
            \frac{1}{d}\mathbb{E}\left[(\beeta^{0})^{\top}\mathbf{g}^{0}(\beeta^{0})\right] = C_{\theta}(0,0)\frac{1}{d}\sum_{i=1}^{n}\mathbb{E}\left[\frac{\partial g_{i}^{0}}{\partial \eta_{i}^{0}}(\beeta^{0})\right]
        \end{equation}
        where $\frac{\partial g_{i}^{0}}{\partial \eta_{i}^{0}}$ denotes the $q \times q$ Jacobian containing the partial derivatives of $g^{0}_{i} : \mathbb{R}^{n \times q} \to \mathbb{R}^{q}$ w.r.t. the line $\eta^{0}_{i}$.
        Since the elements of $\mathbf{v}^{0}$ are i.i.d. sub-gaussian and $d>q$, the matrix $C_{\theta}(0,0)$ is almost surely invertible \cite{vershynin2018high}, therefore 
        \begin{equation}
            \norm{\left(\frac{1}{d}(\mathbf{v}^{0})^{\top}\mathbf{v}^{0}\right)^{-1}\frac{1}{d}(\mathbf{v}^{0}\mathbf{X})^{\top}g^{0}(\mathbf{X}\mathbf{v}^{0})-\frac{1}{d}\sum_{i=1}^{n}\mathbb{E}\left[\frac{\partial g_{i}^{0}}{\partial \eta_{i}^{0}}(\beeta^{0})\right]}_{F} \xrightarrow[n,d \to \infty]{P} 0
        \end{equation}
        which immediately leads to 
        \begin{equation}
            \label{eq:t10}
            \frac{1}{\sqrt{d}}\norm{\mathbf{v}^{0}\left(\frac{1}{d}(\mathbf{v}^{0})^{\top}\mathbf{v}^{0}\right)^{-1}\frac{1}{d}(\mathbf{v}^{0}\mathbf{X})^{\top}g^{0}(\mathbf{X}\mathbf{v}^{0})-\mathbf{v}^{0}\Gamma^{0}}_{F} \xrightarrow[n,d \to \infty]{P} 0
        \end{equation}
        where we introduced the $q \times q$ matrix $\Gamma^{0} = \lim_{n, d \to \infty} \frac{1}{d}\sum_{i=1}^{n}\mathbb{E}\left[\frac{\partial g_{i}^{0}}{\partial \eta_{i}^{0}}(\beeta^{0})\right]$.
        Moving to the third term in Eq.\eqref{eq:decomp_0}, we may use lemma \ref{conv_lemmas_app} to show that, conditionally on $\mathbf{m}^{0}$, 
        \begin{equation}
            \label{eq:t20}
            \frac{1}{\sqrt{d}}\norm{\mathbf{P}^{\perp}_{\mathbf{v}^{0}}\tilde{\mathbf{X}}^{\top}\mathbf{m}^{0}-\tilde{\mathbf{X}}^{\top}\mathbf{m}^{0}}_{F} \xrightarrow[n,d \to \infty]{P} 0
        \end{equation}
        Since $\tilde{\mathbf{X}}$ is independent on $\mathbf{m}^{0}$, we may use lemma \ref{conv_lemmas_app} to show that, conditionally on $\mathbf{m}^{0}$, there exists a $d \times q$ random matrix $\mathbf{u}^{0}$ distributed according to 
        $\mathbf{u}^{0} \sim \mathcal{N}(0,C_{g}(0,0)\otimes \mathbf{I}_{d})$ such that 
        \begin{equation}
            \label{eq:t30}
            \frac{1}{\sqrt{d}}\norm{\tilde{\mathbf{X}}^{\top}\mathbf{m}^{0}-\mathbf{u}^{0}}_{F} \xrightarrow[n,d \to \infty]{P} 0
        \end{equation} 
        where $C_{g}(0,0) = \lim_{n,d \to \infty} \frac{1}{d}(\mathbf{m}^{0})^{\top}\mathbf{m}^{0}$. Finally, note that, by definition of $\mathbf{m}^{0}$, Gaussian concentration of pseudo-Lipschitz functions and Eq.\eqref{inter:eta0}, we have that, with high probability :
        \begin{equation}
            \lim_{n,d \to \infty} \frac{1}{d}(\mathbf{m}^{0})^{\top}\mathbf{m}^{0} = \lim_{n,d \to \infty} \frac{1}{d}\mathbb{E}\left[\mathbf{g}^{0}(\beeta^{0})^{\top}\mathbf{g}^{0}(\beeta^{0})\right].
        \end{equation}
        Combining the results from Eq.\eqref{eq:t10},\eqref{eq:t20},\eqref{eq:t30}. and using the definition of pseudo-Lipschitz functions, we reach that, for any sequence of pseudo-Lipchitz functions of order $k$, $\{\phi_{n}\}_{n\in \mathbb{N}}$:
        \begin{equation}
             \phi_{n}(\mathbf{v}^{1}) \approxP \phi_{n}(\mathbf{h}^{0}(\mathbf{v}^{0})+\mathbf{v}^{0}\Gamma^{0}+\mathbf{u}^{0}) 
        \end{equation}
        which concludes the induction step for $\mathbf{v}^{1}$. \quad \\
        \quad \\
    \paragraph{Induction} 
    Assume that Theorem \ref{th:main_dmft} is verified up to time $t$, i.e. for all iterates up to $\mathbf{r}^{t-1},\mathbf{v}^{t}$. We prove the property for $\mathbf{r}^{t},\mathbf{v}^{t+1}$. \\
    \quad \\
    We shall condition on the $\sigma$-algebra generated by $\mathbf{v}^{0}, ..., \mathbf{v}^{t}, \mathbf{r}^{0}, ..., \mathbf{r}^{t-1}$, denoted $\mathfrak{S}^{t}$. A short induction 
    and application of the Doob-Dynkin lemma show that this $\sigma$-algebra is the same as that generated by $\mathbf{v}^{0},\mathbf{X}^{\top}\mathbf{m}^{0}, ...,\mathbf{X}^{\top}\mathbf{m}^{t-1},\mathbf{X}\mathbf{w}^{0}, ..., \mathbf{X}\mathbf{w}^{t-1}$, where we remind that $\mathbf{w}^{s} = \sum_{k=0}^{s}\mathbf{v}^{k}$ with $\mathbf{w}^{0} = \mathbf{v}^{0}$.
    Let $\mathbf{M}_{t-1},\mathbf{W}_{t-1}$ be the matrices defined by,
    \begin{equation}
    \mathbf{M}_{t-1} = \left[\mathbf{m}^{0} \vert \mathbf{m}^{1} \vert ... \vert \mathbf{m}^{t-1} \right], \mathbf{W}_{t-1} = \left[\mathbf{w}^{0} \vert \mathbf{w}^{1} \vert ... \vert \mathbf{w}^{t-1}\right]
    \end{equation}
    Starting with $\mathbf{r}^{t}$, we may write 
    \begin{align}
        \mathbf{r}^{t}\vert_{\mathfrak{S}^{t}} &= \left(\mathbf{X}\sum_{k=0}^{t}\mathbf{v}^{k}\right)\vert_{\mathfrak{S}^{t}} \\
        &\stackrel{d}= \mathbf{r}_{t-1}+\mathbf{X}\vert_{\mathfrak{S}^{t}}\mathbf{v}^{t} \\
        &\stackrel{d}= \mathbf{r}^{t-1}+\left(\mathbf{P}_{\mathbf{M}_{t-1}}\mathbf{X}+\mathbf{X}\mathbf{P}_{\mathbf{W}_{t-1}}-\mathbf{P}_{\mathbf{M}_{t-1}}\mathbf{X}\mathbf{P}_{\mathbf{W}_{t-1}}+\mathbf{P}^{\perp}_{\mathbf{M}_{t-1}}\tilde{\mathbf{X}}\mathbf{P}^{\perp}_{\mathbf{W}_{t-1}}\right)\mathbf{v}^{t} \\
        &= \mathbf{r}^{t-1}+\left(\mathbf{X}\mathbf{P}_{\mathbf{W}_{t-1}}+\mathbf{P}_{\mathbf{M}_{t-1}}\mathbf{X}\mathbf{P}^{\perp}_{\mathbf{W}_{t-1}}+\mathbf{P}^{\perp}_{\mathbf{M}_{t-1}}\tilde{\mathbf{X}}\mathbf{P}^{\perp}_{\mathbf{W}_{t-1}}\right)\mathbf{v}^{t} \label{eq:inter2}
    \end{align}
    where $\tilde{\mathbf{X}}$ is a copy of $\mathbf{X}$ whose elements are independent of $\mathfrak{S}^{t}$ and $\mathbf{X}$. \newline
    At this point, we introduce an assumption guaranteeing that the projectors are well-defined, in similar fashion to \cite{berthier2020state, gerbelot2021graph}. It will be relaxed at the end of the proof, in Appendix \ref{app_sub:relax}.
\paragraph{Non-degeneracy assumption}
We say that the iteration Eq.(\ref{eq:the_dynamics1}-\ref{eq:the_dynamics2}) satisfies the non-degeneracy assumption if :
\begin{itemize}
    \item almost surely, for all $t$ and all $n \geqslant t$, $\mathbf{M}_{t-1}, \mathbf{W}_{t-1}$ have full column rank.
    \item for all $t $, there exists some constants $c_{M,t},c_{W,t}>0$---independent of n or d---such that almost surely, there exists $n_{0}$ (random) such that, for $n \geqslant n_{0}$, $\sigma_{\min}(\mathbf{M}_{t-1})/\sqrt{n}\geqslant c_{M,t} >0$ and $\sigma_{\min}(\mathbf{W}_{t-1})/\sqrt{n}\geqslant c_{W,t} >0$. \newline
\end{itemize}
    We now turn to the term by term study of Eq.\eqref{eq:inter2}. the first term $\mathbf{r}^{t-1}$ is straightforward to analyze by an immediate use of the induction hypothesis:
    \begin{equation}
        \frac{1}{\sqrt{d}}\norm{\mathbf{r}^{t-1}-\left(\sum_{l=0}^{t-2}\mathbf{g}^{l}(\beeta^{l})R_{\theta}(t-1,l)+\boldsymbol{\omega}^{t-1}\right)}_{F} \xrightarrow[n,d \to \infty]{P} 0
    \end{equation}
    The second term then reads 
    \begin{align}
        \mathbf{X}\mathbf{P}_{\mathbf{W}_{t-1}}\mathbf{v}^{t} &= \mathbf{X}\mathbf{W}_{t-1}\left(\mathbf{W}_{t-1}^{\top}\mathbf{W}_{t-1}\right)^{-1}\mathbf{W}_{t-1}^{\top}\mathbf{v}^{t} \\
        &= \left[\mathbf{r}^{0} \vert \mathbf{r}^{1} \vert ... \vert \mathbf{r}^{t-1}\right] \boldsymbol{\alpha}^{t} \\
        &= \sum_{k=0}^{t-1}\mathbf{r}^{k}\alpha_{k}^{t}
    \end{align}
    where we introduced
    \begin{align}
        \boldsymbol{\alpha}^{t} &= \left(\mathbf{W}_{t-1}^{\top}\mathbf{W}_{t-1}\right)^{-1}\mathbf{W}_{t-1}^{\top}\mathbf{v}^{t} \in\mathbb{R}^{tq \times q} \notag \\
        &= \left(\frac{1}{d}\mathbf{W}_{t-1}^{\top}\mathbf{W}_{t-1}\right)^{-1}\frac{1}{d}\mathbf{W}_{t-1}^{\top}\mathbf{v}^{t},
    \end{align}
    and, for any $0 \leqslant k \leqslant t$, $\alpha_{k}^{t,*}$ denotes the $k-th$, $q \times q$ block of $\boldsymbol{\alpha}^{t,*}$.
    Owing to the non-degeneracy assumption, induction hypothesis and lemma \ref{lemma:pseudo-lip-conv}, the quantity $\frac{1}{d}\mathbf{W}_{t-1}^{\top}\mathbf{W}_{t-1}$ has bounded norm with high probability and converges with high-probability to a deterministic, full-rank $tq \times tq$ matrix. Also, the induction hypothesis and lemma \ref{lemma:pseudo-lip-conv} show that $\frac{1}{d}\mathbf{W}_{t-1}^{\top}\mathbf{v}^{t}$ has bounded norm with high probability and converges with high probability to a deterministic $tq \times q$ matrix. We deduce that $\boldsymbol{\alpha}^{t}$ converges to a deterministic limit $\boldsymbol{\alpha}^{t,*} \in \mathbb{R}^{tq \times q}$ representing 
    the coefficients of the projection of the columns of $\mathbf{v}^{t}$ onto the subspace spanned by the columns of $\mathbf{W}_{t-1}$. Now, let $\bTheta_{t-1} = \left[\btheta^{0} \vert \btheta^{1} ... \vert \btheta^{t-1}\right]$ be the $d \times tq$ matrix whose columns contain the 
    $\boldsymbol{\theta}_{k}$ from Theorem \ref{th:main_dmft} up to time $t-1$. Note that, the induction hypothesis implies that $\boldsymbol{\alpha}^{t,*}$ can also be written as the following limit
    \begin{align}
        \label{eq:alt_proj1}
        &\boldsymbol{\alpha}^{t,*} = \lim_{n,d \to \infty} \mathbb{E}\left[\left(\frac{1}{d}\bTheta^{\top}_{t-1}\bTheta_{t-1}\right)^{-1}\frac{1}{d}\bTheta_{t-1}^{\top}\left(\btheta^{t}-\btheta^{t-1}\right)\right],
    \end{align}
    where the symbol $\approxP$ is to be understood elementwise. This identity will be useful later on. 
    We may then write, using the triangle inequality, 
    \begin{align}
        &\frac{1}{\sqrt{d}}\norm{\mathbf{X}\mathbf{P}_{\mathbf{W}_{t-1}}\mathbf{v}^{t}-\sum_{k=0}^{t-1}\boldsymbol{\eta}^{k}\alpha^{t,*}_{k}}_{F} = \frac{1}{\sqrt{d}}\norm{\sum_{k=0}^{t-1}\mathbf{r}^{k}\alpha^{t}_{k}-\sum_{k=0}^{t-1}\boldsymbol{\eta}^{k}\alpha^{t,*}_{k}}_{F} \\
        & \leqslant \frac{1}{\sqrt{d}}\norm{\sum_{k=0}^{t-1}(\mathbf{r}^{t}-\boldsymbol{\eta}^{t})\alpha_{k}^{t}}_{F}+\frac{1}{\sqrt{d}}\norm{\sum_{k=0}^{t-1}\boldsymbol{\eta}^{k}(\alpha_{k}^{t}-\alpha_{k}^{t,*})}_{F} \\
        & \leqslant \sup_{0 \leqslant k \leqslant t} \norm{\alpha_{k}^{t}}_{op}\sum_{k=0}^{t-1}\frac{1}{\sqrt{d}}\norm{\mathbf{r}^{k}-\beeta^{k}}_{F}+\sum_{k=0}^{t-1}\frac{1}{\sqrt{d}}\norm{\beeta^{k}}_{F}\norm{\alpha^{t}_{k}-\alpha^{t,*}_{k}}_{F}
    \end{align}
 by the induction hypothesis and the non-degeneracy assumption, the quantities $\sup_{0 \leqslant k \leqslant t} \norm{\alpha_{k}^{t}}_{op}$ and $\frac{1}{\sqrt{d}}\norm{\beeta^{t}}_{F}$ are bounded with high probability as $n,d$ go to infinity. Furthermore, the induction hypothesis implies that for any $0 \leqslant k \leqslant t-1$, with high probability
 \begin{align}
    \lim_{n,d \to \infty} \frac{1}{\sqrt{d}}\norm{\mathbf{r}^{k}-\beeta^{k}}_{F} = 0 \quad \mbox{and} \quad \lim_{n,d \to \infty} \norm{\alpha^{t}_{k}-\alpha^{t,*}_{k}}_{F} = 0
 \end{align} 
 which leads to 
 \begin{equation}
    \label{eq:ind_r_1}
    \frac{1}{\sqrt{d}}\norm{\mathbf{X}\mathbf{P}_{\mathbf{W}_{t-1}}\mathbf{v}^{t}-\sum_{k=0}^{t-1}\boldsymbol{\eta}^{k}\alpha^{t,*}_{k}}_{F} \xrightarrow[n,d \to \infty]{P} 0
 \end{equation}
 \\
    \quad \\
    Moving to the third term, we have
    \begin{align}
        \mathbf{P}_{\mathbf{M}_{t-1}}\mathbf{X}\mathbf{P}^{\perp}_{\mathbf{W}_{t-1}}\mathbf{v}^{t} &= \mathbf{M}_{t-1}\left(\mathbf{M}_{t-1}^{\top}\mathbf{M}_{t-1}\right)^{-1}\mathbf{M}_{t-1}^{\top}\mathbf{X}\mathbf{P}_{\mathbf{W}_{t-1}}^{\perp}\mathbf{v}^{t} \\
        &=\mathbf{M}_{t-1}\left(\frac{1}{d}\mathbf{M}_{t-1}^{\top}\mathbf{M}_{t-1}\right)^{-1}\frac{1}{d}\mathbf{M}_{t-1}^{\top}\mathbf{X}\mathbf{P}_{\mathbf{W}_{t-1}}^{\perp}\mathbf{v}^{t}
    \end{align}
    where, using the definition of iteration Eq.~(\ref{eq:the_dynamics1}-\ref{eq:the_dynamics2}) and expanding the projector $\mathbf{P}_{\mathbf{W}_{t-1}}^{\perp} = \mathbf{I}_{d}-\mathbf{P}_{\mathbf{W}_{t-1}}$, we may write
    \begin{align}
        &\frac{1}{d}\mathbf{M}_{t-1}^{\top}\mathbf{X}\mathbf{P}_{\mathbf{W}_{t-1}}^{\perp}\mathbf{v}^{t} = \frac{1}{d}\left[\mathbf{v}^{1}-\bm{h}^{0}(\mathbf{v}^{0}) \vert ...\vert \mathbf{v}^{t}-\bm{h}^{t-1}(\left\{\mathbf{v}^{k}\right\}_{k=0}^{t-1})\right]^{\top}\mathbf{v}^{t} \notag \\
        &\hspace{2.5cm}-\frac{1}{d}\left[\mathbf{v}^{1}-\bm{h}^{0}(\mathbf{v}^{0}) \vert ...\vert \mathbf{v}^{t}-\bm{h}^{t-1}(\left\{\mathbf{v}^{k}\right\}_{k=0}^{t-1})\right]^{\top}\mathbf{P}_{\mathbf{W}_{t-1}}\mathbf{v}^{t}.
    \end{align}
    Now,
    \begin{align}
        &\frac{1}{d}\left[\mathbf{v}^{1}-\bm{h}^{0}(\mathbf{v}^{0}) \vert ...\vert \mathbf{v}^{t}-\bm{h}^{t-1}(\left\{\mathbf{v}^{k}\right\}_{k=0}^{t-1})\right]^{\top}\mathbf{P}_{\mathbf{W}_{t-1}}\mathbf{v}^{t} =  \\
        &=\frac{1}{d}\left[\mathbf{v}^{1}-\bm{h}^{0}(\mathbf{v}^{0}) \vert ...\vert \mathbf{v}^{t}-\bm{h}^{t-1}(\left\{\mathbf{v}^{k}\right\}_{k=0}^{t-1})\right]^{\top}\mathbf{W}_{t-1}\left(\frac{1}{d}\mathbf{W}_{t-1}^{\top}\mathbf{W}_{t-1}\right)^{-1}\frac{1}{d}\mathbf{W}_{t-1}^{\top}\mathbf{v}^{t}
    \end{align}
    Using the induction hypothesis and pseudo-Lipschitz convergence lemma \ref{lemma:pseudo-lip-conv},
    \begin{align}
        &\frac{1}{d}\left[\mathbf{v}^{1}-\bm{h}^{0}(\mathbf{w}^{0}) \vert ...\vert \mathbf{v}^{t}-\bm{h}^{t-1}(\left\{\mathbf{v}^{k}\right\}_{k=0}^{t-1})\right]^{\top}\mathbf{W}_{t-1} \approxP \notag\\
        &\frac{1}{d}\left[\Gamma^{0}\btheta^{0}+\bm{u}^{0} \vert ... \vert \Gamma^{t-1}\btheta^{t-1}+\sum_{k=0}^{t-2}\btheta^{k}R_{l}(t-1,k)+\mathbf{u}^{t-1}\right]^{\top}\bTheta_{t-1} \\
        &=\underbrace{\frac{1}{d}\left[\Gamma^{0}\btheta^{0}\vert ... \vert \Gamma^{t-1}\btheta^{t-1}+\sum_{k=0}^{t-2}\btheta^{k}R_{l}(t-1,k)\right]^{\top}}_{\in \thickspace \mbox{span}(\bTheta_{t-1})}\bTheta_{t-1}+\frac{1}{d}\left[\bm{u}^{0} \vert ... \vert \bm{u}^{t-1}\right]^{\top}\bTheta_{t-1}
    \end{align}
    and 
    \begin{equation}
        \frac{1}{d}\mathbf{W}_{t-1}^{\top}\mathbf{v}^{t} \approxP \frac{1}{d}\bTheta_{t-1}^{\top}\left(\btheta^{t}-\btheta^{t-1}\right),
    \end{equation}
    where we also have 
    \begin{align}
        \frac{1}{d}\mathbf{W}_{t-1}^{\top}\mathbf{W}_{t-1} \approxP \frac{1}{d}\bTheta_{t-1}^{\top}\bTheta_{t-1},
    \end{align}
    the limit of which is an invertible matrix owing to the non-degeneracy assumption.
We thus reach 
\begin{align}
    &\frac{1}{d}\left[\mathbf{v}^{1}-\bm{h}^{0}(\mathbf{w}^{0}) \vert ...\vert \mathbf{v}^{t}-\bm{h}^{t-1}(\mathbf{w}^{t-1})\right]^{\top}\mathbf{v}^{t} \approxP \notag \\
    &\hspace{2.5cm}\frac{1}{d}\left[\bGamma^{0}\btheta^{0}+\bm{u}^{0} \vert ... \vert \bGamma^{t-1}\btheta^{t-1}+\sum_{k=0}^{t-2}\btheta^{k}R_{l}(t-1,k)+\bm{u}^{t-1}\right]^{\top}\left(\btheta^{t}-\btheta^{t-1}\right) \\
    &\mbox{and} \quad \frac{1}{d}\left[\mathbf{v}^{1}-\bm{h}^{0}(\mathbf{w}^{0}) \vert ...\vert \mathbf{v}^{t}-\bm{h}^{t-1}(\mathbf{w}^{t-1})\right]^{\top}\mathbf{P}_{\mathbf{W}_{t-1}}\mathbf{v}^{t} \approxP \notag \\
    &\hspace{2.5cm}\frac{1}{d}\left[\bGamma^{0}\btheta^{0}+\bm{u}^{0} \vert ... \vert \bGamma^{t-1}\btheta^{t-1}+\sum_{k=0}^{t-2}\btheta^{k}R_{l}(t-1,k)+\bm{u}^{t-1}\right]^{\top} \\
    &\hspace{6.5cm}\bTheta_{t-1}\left(\frac{1}{d}\bTheta_{t-1}^{\top}\bTheta_{t-1}\right)^{-1}\frac{1}{d}\bTheta_{t-1}^{\top}\left(\btheta^{t}-\btheta^{t-1}\right) 
\end{align}
which, when combined, leads to 
    \begin{align}
        \frac{1}{d}\mathbf{M}_{t-1}^{\top}\mathbf{X}\mathbf{P}_{\mathbf{W}_{t-1}}^{\perp}\mathbf{v}^{t} &\approxP \frac{1}{d}\left[\bm{u}^{0} \vert ... \vert \bm{u}^{t-1}\right]^{\top}\left(\btheta^{t}-\btheta^{t-1}\right)-\frac{1}{d}\left[\bm{u}^{0} \vert ... \vert \bm{u}^{t-1}\right]^{\top}\mathbf{P}_{\bTheta_{t-1}}\left(\btheta^{t}-\btheta^{t-1}\right) \\
        &\approxP \frac{1}{d}\mathbb{E}\left[\left[\bm{u}^{0} \vert ... \vert \bm{u}^{t-1}\right]^{\top}\left(\btheta^{t}-\btheta^{t-1}\right)\right]-\frac{1}{d}\mathbb{E}\left[\left[\bm{u}^{0} \vert ... \vert \bm{u}^{t-1}\right]^{\top}\bTheta_{t-1}\right]\boldsymbol{\alpha}^{t,*},
    \end{align}
    where we used the expression for $\boldsymbol{\alpha}^{t,*}$ given by Eq.\eqref{eq:alt_proj1} in the last line.
    Now, remembering the equation defining $\btheta^{s}$ for any $0\leqslant s \leqslant t$ in Theorem \ref{th:main_dmft}, we may use Stein's lemma \ref{matrix-stein} to obtain 
    \begin{align}
        \forall \thickspace 0\leqslant r,s \leqslant t \quad \frac{1}{d}(\bm{u}^{r})^{\top}\btheta^{s}(\bm{u}^{0},\bm{u}^{1},...,\bm{u}^{s-1}) &\approxP \frac{1}{d}\sum_{i=0}^{s-1}C_{g}(i,r)\sum_{j=1}^{d}\mathbb{E}\left[\frac{\partial \theta_{j}^{s}}{\partial u^{i}_{j}}\right] \notag \\
        &\approxP \sum_{i=0}^{s-1}C_{g}(i,r)R_{\theta}(s,i)
    \end{align}
    where $\boldsymbol{\theta}^{s}$ is considered a function with domain $\mathbb{R}^{d \times tq}$ and image in $\mathbb{R}^{d\times q}$, taking the $\{\mathbf{u}^{k}\}_{0 \leqslant k \leqslant t-1}$. The notation $\frac{\partial \theta_{j}^{s}}{\partial u^{i}_{j}}$ then denotes the $q \times q$
    jacobian matrix obtained with the partial derivatives of $\theta^{s}_{j}$, the restriction of $\btheta^{s}$ to the $j-th$ line of its image, viewed as a function going from $\mathbb{R}^{d \times tq}$ to $\mathbb{R}^{q}$,  with respect to the $j-th$ line of $\mathbf{u}^{i}$.
    Letting $\mathbf{C}_{g,t}$ be the $tq\times tq$ covariance matrix of the lines of $\left[\bm{u}^{0} \vert ... \vert \bm{u}^{t-1}\right] \in \mathbb{R}^{d \times tq}$ for any $t$, we can now write 
    \begin{align}
        &\frac{1}{d}\left[\bm{u}^{0} \vert ... \vert \bm{u}^{t-1}\right]^{\top}\btheta^{t} \approxP \mathbf{C}_{g,t}\begin{bmatrix}\frac{1}{d}\sum_{j=1}^{d}\mathbb{E}\left[\frac{\partial \theta_{j}^{t}}{\partial u^{0}_{j}}\right] \\...\\\frac{1}{d}\sum_{j=1}^{d}\mathbb{E}\left[\frac{\partial \theta_{j}^{t}}{\partial u^{t-1}_{j}}\right]\end{bmatrix} \\
        &= \mathbf{C}_{g,t}\begin{bmatrix}R_{\theta}(t,0) \\... \\R_{\theta}(t,t-1)\end{bmatrix} = \mathbf{C}_{g,t}\mathbf{R}_{\theta,t}
    \end{align}
    where we defined the $tq \times q$ matrix $\mathbf{R}_{\theta,t} = \begin{bmatrix}R_{\theta}(t,0) \\... \\R_{\theta}(t,t-1)\end{bmatrix}$. \\
    Similarly, for any $0 \leqslant s \leqslant t$
    \begin{align}
        \frac{1}{d}\left[\bm{u}^{0} \vert ... \vert \bm{u}^{t-1}\right]^{\top}\btheta^{s} \approxP \mathbf{C}_{g,t}\begin{bmatrix}\frac{1}{d}\sum_{j=1}^{d}\mathbb{E}\left[\frac{\partial \theta_{j}^{s}}{\partial u^{0}_{j}}\right] \\... \\\frac{1}{d}\sum_{j=1}^{d}\mathbb{E}\left[\frac{\partial \theta_{j}^{s}}{\partial u^{s-1}_{j}}\right] \\0 \\... \\0\end{bmatrix} = \mathbf{C}_{g,t}\mathbf{R}_{\theta,s}
    \end{align}
    where the zeroes come from the fact that $\theta^{s}$ is not an algebraic function of the $u^{l}$ for $l \geqslant s$, which is coherent with the causality from the physics approach, even though the 
    Gaussian process $u^{l}$ is correlated across all $0\leqslant l \leqslant t-1$. Note that the matrices $\mathbf{R}_{\theta,s}$ are defined in such a way that, for any $0 \leqslant s \leqslant t$, $\mathbf{R}_{\theta,s}$ all have the same dimension $tq \times q$. We thus reach
    \begin{equation}
        \frac{1}{d}\mathbf{M}_{t-1}^{\top}\mathbf{X}\mathbf{P}_{\mathbf{W}_{t-1}}\mathbf{v}^{t} \approxP \mathbf{C}_{g,t}\left(\mathbf{R}_{\theta,t}-\mathbf{R}_{\theta,t-1}-\left[\mathbf{R}_{\theta,0}\vert \mathbf{R}_{\theta,1} \vert ... \vert \mathbf{R}_{\theta,t-1}\right]\boldsymbol{\alpha}^{t,*}\right)
    \end{equation}
    Also, due to the induction hypothesis
    \begin{align}
        \label{eq:inter_proj1}
        \frac{1}{d}\mathbf{M}_{t-1}^{\top}\mathbf{M}_{t-1} \approxP \mathbf{C}_{g,t}.
    \end{align}
    Combining the above two lines, we obtain 
    \begin{align}
        (\frac{1}{d}\mathbf{M}_{t-1}^{\top}\mathbf{M}_{t-1})^{-1}\frac{1}{d}\mathbf{M}_{t-1}^{\top}\mathbf{X}\mathbf{P}_{\mathbf{W}_{t-1}}\mathbf{v}^{t} \approxP \left(\mathbf{R}_{\theta,t}-\mathbf{R}_{\theta,t-1}-\left[\mathbf{R}_{\theta,0}\vert \mathbf{R}_{\theta,1} \vert ... \vert \mathbf{R}_{\theta,t-1}\right]\boldsymbol{\alpha}^{t,*}\right)
    \end{align}
    which, along with the induction hypothesis ensuring that $\frac{1}{\sqrt{d}}\norm{\mathbf{M}_{t-1}}_{F}$ is bounded with high probability, implies that 
    \begin{align}
        &\frac{1}{\sqrt{d}}\norm{\mathbf{M}_{t-1}(\mathbf{M}_{t-1}^{\top}\mathbf{M}_{t-1})^{-1}\mathbf{M}_{t-1}^{\top}\mathbf{X}\mathbf{P}_{\mathbf{W}_{t-1}}\mathbf{v}^{t} - \mathbf{M}_{t-1}\left(\mathbf{R}_{\theta,t}-\mathbf{R}_{\theta,t-1}-\left[\mathbf{R}_{\theta,0}\vert \mathbf{R}_{\theta,1} \vert ... \vert \mathbf{R}_{\theta,t-1}\right]\boldsymbol{\alpha}^{t,*}\right)}_{F} \\
        &\xrightarrow[n,d \to \infty]{P} 0 \notag 
    \end{align}
    Now, for any $0 \leqslant s \leqslant t$, by definition of $\mathbf{R}_{\theta,s}$, we have that
    \begin{equation}
        \mathbf{M}_{t-1}\mathbf{R}_{\theta,s} = \sum_{l=0}^{t-1}\mathbf{m}^{l}R_{\theta}(s,l).
    \end{equation}
    The triangle inequality then gives
    \begin{align}
        \frac{1}{\sqrt{d}}\norm{\sum_{l=0}^{t-1}\mathbf{m}^{l}R_{\theta}(s,l)-\sum_{l=0}^{t-1}\mathbf{g}^{l}(\eta^{l})R_{\theta}(s,l)}_{F} \leqslant \sup_{0 \leqslant l \leqslant t-1}\norm{R_{\theta}(s,l)}_{op} \sum_{l=0}^{t-1}\frac{1}{\sqrt{d}}\norm{\mathbf{m}^{l}-\mathbf{g}^{l}(\boldsymbol{\eta}^{l})}_{F}
    \end{align}
    The induction hypothesis then shows that, for any $0 \leqslant l \leqslant t-1$, $\frac{1}{\sqrt{d}}\norm{\mathbf{m}^{l}-\mathbf{g}^{l}(\beeta^{l})}_{F}$ goes to zero with high probability 
    when $n,d \to \infty$, and $\sup_{0 \leqslant l \leqslant t-1}\norm{R_{\theta}(s,l)}_{op}$ is bounded. Thus 
    \begin{equation}
        \frac{1}{\sqrt{d}}\norm{\mathbf{M}_{t-1}\mathbf{R}_{\theta,s}-\sum_{l=0}^{t-1}\mathbf{g}^{l}(\beeta^{l})R_{\theta}(s,l)}_{F} \xrightarrow[n,d \to \infty]{P} 0,
    \end{equation}
    for any $0 \leqslant s \leqslant t$, and where we remind that $R_{\theta}(s,l) = 0_{q \times q}$ for any $l>s$. In particular, we have that 
    \begin{equation}
        \frac{1}{\sqrt{d}}\norm{\mathbf{M}_{t-1}\left[\mathbf{R}_{\theta,0}\vert \mathbf{R}_{\theta,1} \vert ... \vert \mathbf{R}_{\theta,t-1}\right]\boldsymbol{\alpha}^{t,*}-\sum_{k=0}^{t-1}\left(\sum_{l'=0}^{k}\bm{g}^{l'}(\beeta^{l'})R_{\theta}(k,l')\right)\alpha^{t,*}_{k}}_{F} \xrightarrow[n,d \to \infty]{P} 0
    \end{equation}
    Moving to the fourth term in Eq.\eqref{eq:inter2}, the independence $\tilde{\mathbf{X}}$ on $\mathfrak{S}^{t}$ and lemma \ref{conv_lemmas_app} show that, with high probability 
    \begin{equation}
       \lim_{n,d \to \infty} \frac{1}{\sqrt{d}}\norm{\mathbf{P}^{\perp}_{\mathbf{M}_{t-1}}\tilde{\mathbf{X}}\mathbf{P}^{\perp}_{\mathbf{W}_{t-1}}\mathbf{v}^{t}-\tilde{\mathbf{X}}\mathbf{P}^{\perp}_{\mathbf{W}_{t-1}}\mathbf{v}^{t}}_{F} = 0
    \end{equation}
    Furthermore, using the induction hypothesis, lemma \ref{lemma:pseudo-lip-conv} and lemma \ref{conv_lemmas_app}, there exists a $n \times q$ random matrix $\tilde{\bomega}^{t} \sim \mathcal{N}(0,\mathbf{C}_{v,t}^{\perp}\otimes \mathbf{I}_{n})$ such that 
    \begin{equation}
        \frac{1}{\sqrt{d}}\norm{\tilde{\mathbf{X}}\mathbf{P}_{\mathbf{W}_{t-1}}^{\perp}\mathbf{v}^{t}-\tilde{\bomega}^{t}}_{F} \xrightarrow[n,d \to \infty]{P} 0,
    \end{equation}
    where $\mathbf{C}_{v,t}^{\perp} = \lim_{d \to \infty} \frac{1}{d} \left(\mathbf{P}_{\mathbf{W}_{t-1}}^{\perp}\mathbf{v}^{t}\right)^{\top}\left(\mathbf{P}_{\mathbf{W}_{t-1}}^{\perp}\mathbf{v}^{t}\right)$. Coming back to Eq.\eqref{eq:inter2}, 
    we can now combine the results obtained above with the triangle inequality to obtain the following asymptotic representation of $\mathbf{r}^{t}\vert_{\mathfrak{S}^{t}}$ : 
    \begin{align}
        &\frac{1}{\sqrt{d}}\vert\vert\mathbf{r}^{t}\vert_{\mathfrak{S}^{t}}-\bigg(\sum_{l=0}^{t-2}\bm{g}^{l}(\beeta^{l})R_{\theta}(t-1,l)+\bomega^{t-1}+\sum_{k=0}^{t-1}\left(\sum_{l'=0}^{k}\bm{g}^{l'}(\beeta^{l'})R_{\theta}(k,l')+\bomega^{k}\right)\alpha^{t,*}_{k}  \notag \\
        &+\sum_{l=0}^{t-1}\mathbf{g}^{l}(\beeta^{l})R_{\theta}(t,l)-\sum_{l=0}^{t-2}\mathbf{g}^{l}(\beeta^{l})R_{\theta}(t-1,l)-\sum_{k=0}^{t-1}\left(\sum_{l'=0}^{k}\bm{g}^{l'}(\beeta^{l'})R_{\theta}(k,l')\right)\alpha^{t,*}_{k}+\tilde{\bomega}^{t} \bigg)\vert \vert_{F} \notag \\
        &\xrightarrow[n,d \to \infty]{P} 0
    \end{align}
    In the above expression, all the terms of the form $\sum_{l'=0}^{k}\bm{g}^{l'}(\beeta^{l'})R_{\theta}(k,l')\alpha^{t,*}_{k}$ for $0 \leqslant k \leqslant t-2$ simplify, leading to
    \begin{align}
        \frac{1}{\sqrt{d}}\norm{\mathbf{r}^{t}\vert_{\mathfrak{S}^{t}}-\sum_{k=0}^{t-1}\bm{g}^{k}(\beeta^{k})R_{\theta}(t,k)+\sum_{k=0}^{t-1}\bomega^{k}\alpha_{k}^{t,*}+\bomega^{t-1}+\tilde{\bomega}^{t}}_{F} \xrightarrow[n,d \to \infty]{P} 0.
        \label{eq:conv1}
    \end{align}
     Now, consider a sequence $\{\phi_{n}\}_{n \in \mathbb{N}}$ of pseudo-Lipschitz functions of order $k$. Then,
    \begin{align}
        \phi_{n}\left(\mathbf{r}^{0},\mathbf{r}^{1},...,\mathbf{r}^{t}\right)\vert_{\mathfrak{S}^{t}} \stackrel{d}{=} \phi_{n}\left(\mathbf{r}^{0},\mathbf{r}^{1},...,\mathbf{r}^{t}\vert_{\mathfrak{S}^{t}}\right)
    \end{align}
    and there exists a constant $L$ independent on $n,d$ such that 
    \begin{align}
        &\abs{\phi_{n}\left(\mathbf{r}^{0},\mathbf{r}^{1},...,\mathbf{r}^{t}\right)\vert_{\mathfrak{S}^{t}}-\phi_{n}\left(\mathbf{r}^{0},\mathbf{r}^{1},...,\sum_{k=0}^{t-1}\bm{g}^{k}(\beeta^{k})R_{\theta}(t,k)+\sum_{k=0}^{t-1}\bomega^{k}\alpha_{k}^{t,*}+\bomega^{t-1}+\tilde{\bomega}^{t}\right)} \leqslant \\
        &L\left(1+\frac{1}{\sqrt{d}}\sum_{k=0}^{t-1}\norm{\mathbf{r}^{k}}_{F}+\frac{1}{\sqrt{d}}\norm{\mathbf{r}_{t}\vert_{\mathfrak{S}^{t}}}_{F}+\frac{1}{\sqrt{d}}\norm{\sum_{k=0}^{t-1}\bm{g}^{k}(\beeta^{k})R_{\theta}(t,k)+\sum_{k=0}^{t-1}\bomega^{k}\alpha_{k}^{t,*}+\bomega^{t-1}+\tilde{\bomega}^{t}}_{F}\right) \notag \\
        &\frac{1}{\sqrt{d}}\norm{\mathbf{r}^{t}\vert_{\mathfrak{S}^{t}}-\sum_{k=0}^{t-1}\bm{g}^{k}(\beeta^{k})R_{\theta}(t,k)+\sum_{k=0}^{t-1}\bomega^{k}\alpha_{k}^{t,*}+\bomega^{t-1}+\tilde{\bomega}^{t}}_{F} 
    \end{align} 
    The induction hypothesis shows that, for any $0 \leqslant k \leqslant t-1$, the quantity $\frac{1}{\sqrt{d}}\norm{\mathbf{r}^{t}}_{F}$ is bounded with high probability. The summability assumptions for the update functions $\mathbf{g}^{k}$ in \ref{main_assum_1}-\ref{main_assum_7} and the definition of the DMFT equations in Theorem \ref{th:main_dmft} ensure that the quantity $\frac{1}{\sqrt{d}}\norm{\sum_{k=0}^{t-1}\bm{g}^{k}(\beeta^{k})R_{\theta}(t,k)+\sum_{k=0}^{t-1}\bomega^{k}\alpha_{k}^{t,*}+\bomega^{t-1}+\tilde{\bomega}^{t}}_{F}$ is bounded with high probability. Finally, the analysis carried out above and Eq.\eqref{eq:conv1} show that $\frac{1}{\sqrt{d}}\norm{\mathbf{r}_{t}\vert_{\mathfrak{S}^{t}}}$ is bounded with high probability and that 
    \begin{align}
        \phi_{n}\left(\mathbf{r}^{0},\mathbf{r}^{1},...,\mathbf{r}^{t}\right)\vert_{\mathfrak{S}^{t}} \approxP \phi_{n}\left(\mathbf{r}^{0},\mathbf{r}^{1},...,\sum_{k=0}^{t-1}\bm{g}^{k}(\beeta^{k})R_{\theta}(t,k)+\sum_{k=0}^{t-1}\bomega^{k}\alpha_{k}^{t,*}+\bomega^{t-1}+\tilde{\bomega}^{t}\right).
    \end{align}
    We thus recover the correct form for the memory term, matching that of $\boldsymbol{\eta}_{t}$ in Theorem \ref{th:main_dmft}.
    To verify that $\sum_{k=0}^{t-1}\bm{g}^{k}(\beeta^{k})R_{\theta}(t,k)+\sum_{k=0}^{t-1}\bomega^{k}\alpha_{k}^{t,*}+\bomega^{t-1}+\tilde{\bomega}^{t}$ has the same distribution as $\boldsymbol{\eta}_{t}$, we are left with checking that the Gaussian process term has the right covariance.
    Define 
    \begin{align}
        \bomega^{t} = \sum_{k=0}^{t-1}\bomega^{k}\alpha_{k}^{t,*}+\bomega^{t-1}+\tilde{\bomega}^{t}.
    \end{align}
    Which is indeed a Gaussian random vector (with elements in $\mathbb{R}^{q}$).
    To check that this is the correct covariance, we start by noticing that, for any $s<t$ Theorem \ref{th:main_dmft} states that:
    \begin{align}
        &\frac{1}{d}(\mathbf{w}^{s})^{\top}\mathbf{w}^{t} = \frac{1}{d}(\mathbf{w}^{s})^{\top}\mathbf{w}^{t-1}+\frac{1}{d}(\mathbf{w}^{s})^{\top}\mathbf{v}^{t} \\
        &\approxP C_{\theta}(s,t-1)+\frac{1}{d}(\mathbf{w}^{s})^{\top}\mathbf{v}^{t}
    \end{align}
    Then, using the induction hypothesis and the fact that $\tilde{\bomega}^{t}$ is independent from any $\bomega^{s}$ for any $s<t$:
    \begin{align}
        \frac{1}{d}\mathbb{E}\left[(\bomega^{s})^{\top}\bomega^{t}\right] &= \frac{1}{d}\sum_{k=0}^{t-1}\mathbb{E}\left[(\bomega^{s})^{\top}\bomega^{s}\right]\alpha_{k}^{t,*}+\frac{1}{d}\mathbb{E}\left[(\bomega^{s})^{\top}\bomega^{t-1}\right] \\
        &=\sum_{k=0}^{t-1}C_{\theta}(s,k)\alpha_{k}^{t,*}+C_{\theta}(s,t-1) \\
        &\approxP \frac{1}{d}(\mathbf{w}^{s})^{\top}\mathbf{W}_{t-1}\left(\mathbf{W}_{t-1}^{\top}\mathbf{W}_{t-1}\right)^{-1}\mathbf{W}_{t-1}^{\top}\mathbf{v}^{t}+C_{\theta}(s,t-1) \\
        & = \frac{1}{d}\left(\mathbf{P}_{\mathbf{W}_{t-1}}\mathbf{w}^{s}\right)^{\top}\mathbf{v}^{t}+C_{\theta}(s,t-1) \\
        &\approxP \frac{1}{d}(\mathbf{w}^{s})^{\top}\mathbf{v}^{t}+C_{\theta}(s,t-1)
    \end{align}
    We then check for $s=t$, noticing that 
    \begin{align}
        \frac{1}{d}(\mathbf{w}^{t})^{\top}\mathbf{w}^{t} &= \frac{1}{d}\left(\mathbf{w}^{t-1}+\mathbf{v}^{t}\right)^{\top}\left(\mathbf{w}^{t-1}+\mathbf{v}^{t}\right) \\
        &\approxP C_{\theta}(t-1,t-1)+\frac{1}{d}(\mathbf{v}^{t})^{\top}\left(\mathbf{w}^{t-1}+\mathbf{v}^{t}\right)
    \end{align}
    \begin{align}
        &\frac{1}{d}\mathbb{E}\left[(\bomega^{t})^{\top}\bomega^{t}\right] = \frac{1}{d}\mathbb{E}\left[\left(\sum_{k=0}^{t-1}\bomega^{k}\alpha_{k}^{t,*}+\bomega^{t-1}+\tilde{\bomega}^{t}\right)^{\top}\left( \sum_{k=0}^{t-1}\bomega^{k}\alpha_{k}^{t,*}+\bomega^{t-1}+\tilde{\bomega}^{t}\right)\right] \\
        &= C_{\theta}(t-1,t-1)+\sum_{k,k'=0}^{t-1}(\alpha_{k'}^{t,*})^{\top}C_{\theta}(k,k')\alpha_{k}^{t,*}+2\sum_{k=0}^{t-1}C_{\theta}(t-1,k)\alpha_{k}^{t,*}+C_{v,t}^{\perp}\\
        &\approxP \frac{1}{d}(\mathbf{w}^{t-1})^{\top}\mathbf{w}^{t-1}+\frac{1}{d}\left(\mathbf{P}_{\mathbf{W}_{t-1}}\mathbf{v}^{t}\right)^{\top}\left(\mathbf{P}_{\mathbf{W}_{t-1}}\mathbf{v}^{t}\right)+\frac{1}{d}\left(\mathbf{P}_{\mathbf{W}_{t-1}}^{\perp}\mathbf{v}^{t}\right)^{\top}\left(\mathbf{P}_{\mathbf{W}_{t-1}}^{\perp}\mathbf{v}^{t}\right) \notag \\
        &+2\frac{1}{d}\mathbf{w}_{t-1}^{\top}\mathbf{v}^{t} \\
        &\approxP \frac{1}{d}\left(\mathbf{w}^{t-1}+\mathbf{v}^{t}\right)^{\top}\left(\mathbf{w}^{t-1}+\mathbf{v}^{t}\right)
    \end{align}
    We thus recover the correct covariance and the statement is proven for $\mathbf{r}^{t}$. \\
\quad \\
The rest of the proof consists in completing the induction on $\mathbf{u}^{t+1}$, in similar fashion to what has been presented for $\bm{r}^{t}$, and relaxing the non-degeneracy assumption using an existing argument from \cite{berthier2020state,gerbelot2021graph}.
The detail is given in Appendix \ref{sec:app_proof}.

\section{Numerical solution of the equations}
\label{sec:numerics}
In this section, we display the numerical solution of the self-consistent DMFT equations in comparison to numerical simulations. We focus on the special case of teacher-student binary classification performed by a single-layer neural network trained via the multi-pass SGD algorithm described in section \ref{sec:algs}. In this setting, for each sample $\mathbf{x}_{\mu}\sim\mathcal{N}(\bm{0}_d, \bm{I}_d)$, $\mu=1,\ldots,n$, the corresponding label is generated by a Gaussian teacher vector $\mathbf{w}^*\in\mathbb{R}^d$, $w^*_i\sim\mathcal{N}(0,\frac 1d)$ i.i.d., as $y_\mu={\rm sign}(\mathbf{x}_\mu^\top\mathbf{w^*})$. Introduced in the seminal work \cite{gardner1989three}, the binary teacher-student perceptron is a widely studied prototype model for classification in the statistical physics literature. Recently, the performance of this model at empirical risk minimization has been put on rigorous ground \cite{aubin2020generalization}. Here, we adopt it as a working example to show the effectiveness of DMFT equations beyond the infinite-dimensional limit. Indeed, we find a good agreement with simulations even at moderately low system size. The learning is performed by a single-layer neural network parametrized by the weight vector $\mathbf{w}\in\mathbb{R}^d$. The empirical risk is given by Eq.~\eqref{eq:ERM_definition}: $\mathcal{L}(\mathbf{w})=\sum_{\mu=1}^n l(\mathbf{x}_\mu^\top\mathbf{w},y_\mu)+F(\mathbf{w})$, with the logistic loss function $l(r,y)=\log(1+{\rm e}^{-yr})$ and ridge regularization $F(\mathbf{w})=\lambda \Vert \mathbf{w}\Vert^2_2/2$ of strength $\lambda\geq 0$. We consider a random initialization $\mathbf{w}^0$ with i.i.d.~standard Gaussian components $w_0^i\sim\mathcal{N}(0,\frac 1d)$, $\forall i=1,\ldots, d$. The high-dimensional SGD dynamics illustrated in the first example of section \ref{sec:main_result}, Eq.~\eqref{eq:sgd_iteration}, is effectively tracked by the DMFT system in corollary~\ref{th:main-dmft-separable}. While it is possible to integrate directly the DMFT system in corollary~\ref{th:main-dmft-separable} with an analogous strategy as the one presented below, it turns out that it is more efficient to integrate a simpler version, that we derive 
in Appendix \ref{app:numerics}. The resulting DMFT equations are
\begin{align}
\label{eq:eff_eta_code}
\begin{split}
    \eta^{t+1}& =(1-\gamma\lambda+\Gamma^t)\eta^t-\frac \gamma b s^t \,l'\left(\eta^t+\eta^*m^t\right)+\sum_{k=0}^{t-1}R_g(t,k)\eta^k+u^t\in\mathbb{R},\\
    m^t&=(1-\gamma\lambda)m^t-\upsilon^t,\\
    R_{g}(t,s) &= -\alpha\gamma \mathbb{E}\left[s^{t}\frac{\partial l^{'}}{\partial \omega^{s}}(\eta^t+\eta^*m^t)\right] \in \mathbb{R}, \\
            \Gamma^{t} &= -\alpha\gamma \mathbb{E}\left[s^{t}l^{''}(\eta^t+\eta^*m^t)\right] \in \mathbb{R},\\
            C_{g}(t,s) &= \alpha\gamma^{2} \mathbb{E}\left[s^{s}s^{t}l^{'}(\eta^s+\eta^*m^s)l^{'}(\eta^t+\eta^*m^t)^{\top}\right] \in \mathbb{R},\\
            \upsilon^t&=\alpha\gamma\mathbb{E}\left[s^tl'(\eta^t+\eta^*m^t)\eta^*\right],
    \end{split}
\end{align}
where $u^t$ is a Gaussian process in $\mathbb{R}$ with covariance $C_g(t,k)$ and the definitions of $C_g(t,k)$, $R_g(t,k)$, $\Gamma^t$ are the same as those in corollary~\ref{th:main-dmft-separable}. We have also performed a translation of the pre-activation effective variable $\eta^t\leftarrow \eta^t-\eta^* m^t$, where $\eta^*$ encodes the effective teacher pre-activation $\bm{X}\bm{w^*}$ and $m^t$ captures the projection of the weight vector onto the teacher ${\bm{w}^t}^\top\bm{w^*}$. Notice that Eq.~\eqref{eq:eff_eta_code} only involves one effective stochastic process and is therefore simpler to iterate.

The numerics proceeds by iterations, starting by a random guess of the memory and response kernels, as well as the auxiliary functions. The DMFT equations are then integrated numerically at fixed kernels and auxiliary functions. The kernels and functions are in turn updated by averaging over the generated stochastic processes. The numerical implementation of this procedure is available at \\
\url{https://github.com/SPOC-group/Rigorous-dynamical-mean-field-theory}. 

This numerical procedure has been first presented in \cite{EO92,EO94}. More recently, it has been adapted further to other applications
(see, e.g., \cite{Roy_2019,MSZ20,mignacco2020dynamical}). 

Once the kernels have reached convergence, we can use their final expressions to sample the stochastic processes for the effective weight $\theta^t$ and pre-activation $\eta^t+\eta^* m^t$, and use them to compute the averages of the quantities of interest, for instance $m^t$ and the weight vector norm $C_\theta(t,t)$. Since we aim at assessing the classification performance, we are interested in the cosine similarity between the weight vector and the signal: $m^t/\sqrt{C_\theta(t,t)}$. Indeed, the generalization error only depends on this quantity \cite{aubin2020generalization}.  

\begin{figure}[t!]
    \centering
    \includegraphics[scale=.47]{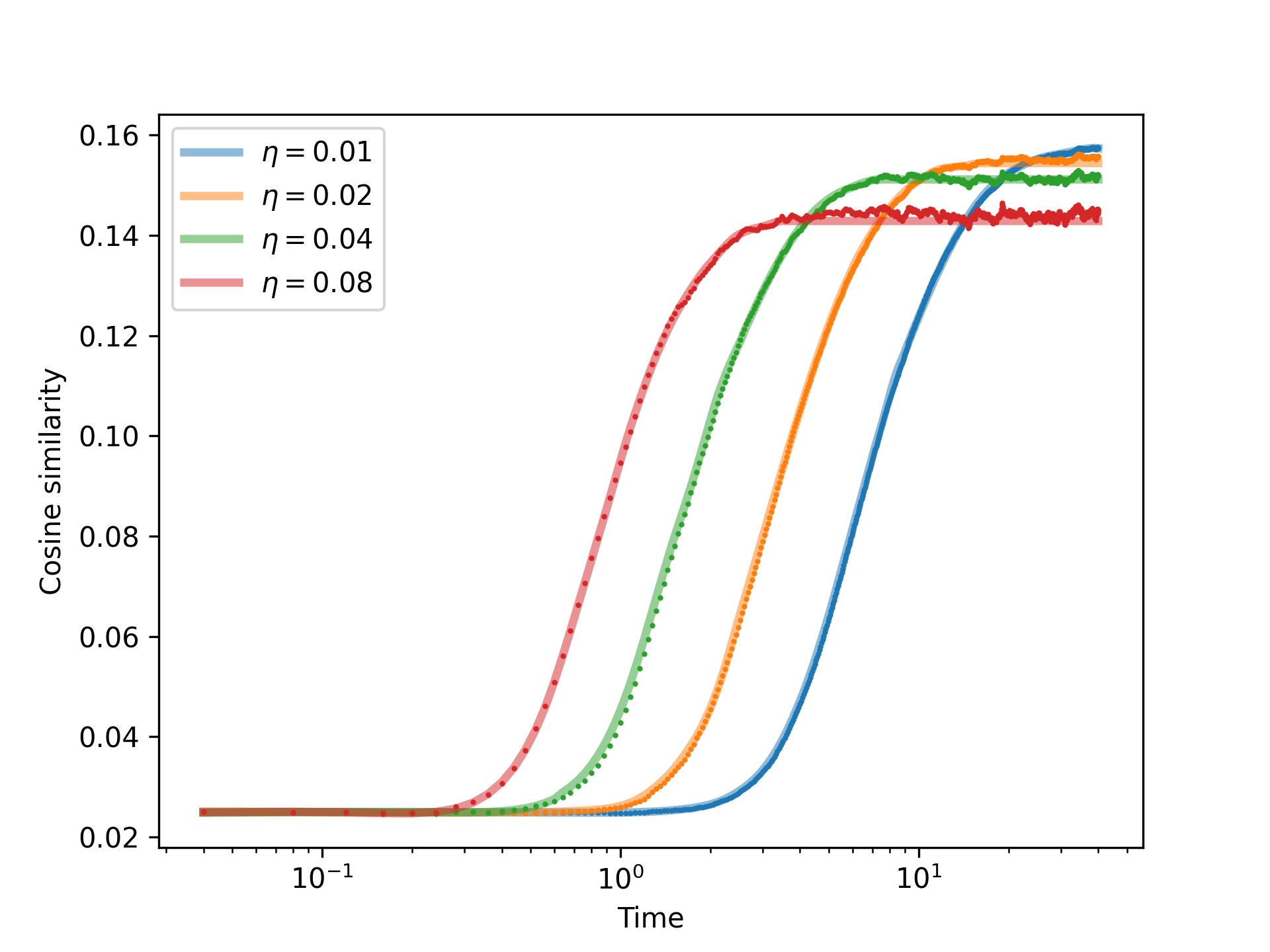}
    \includegraphics[scale=.47]{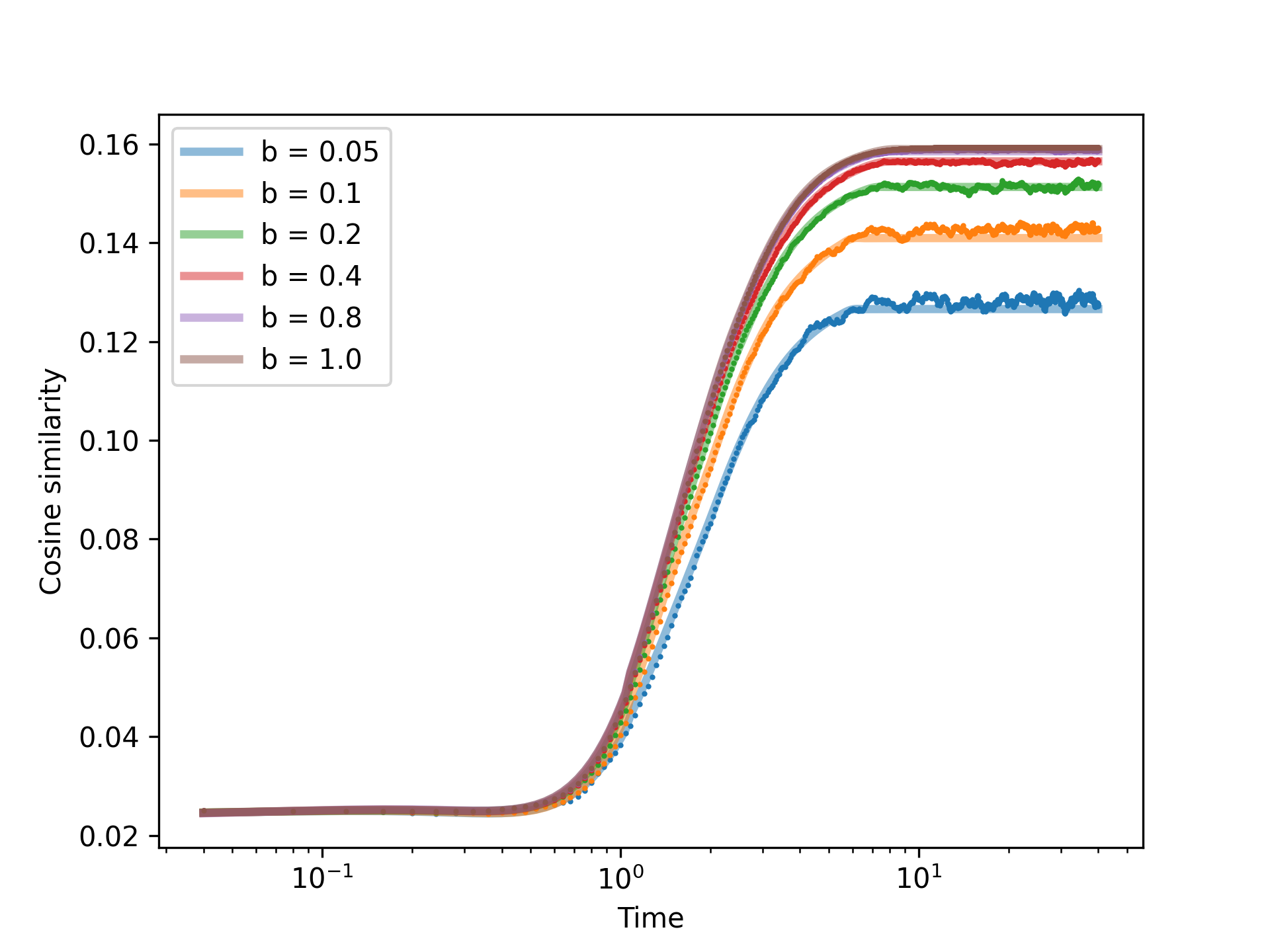}
    \caption{Average cosine similarity with the signal as a function of time for different values of the learning rate $\gamma$ (left panel) and batch size $b$ (right panel). Parameters $\lambda=1$, $\alpha=0.9$, $b=0.2$ on the left, and $\gamma=0.04$ on the right. 
    Continuous pale lines: solution of the DMFT equations in the high-dimensional limit. Dots: simulations with $d=1000$. On the left: different colors indicate different learning rates $\gamma$ with $b=0.2$. 
    }
    \label{fig:DMFT_vs_simulations}
\end{figure}

In Fig.~\ref{fig:DMFT_vs_simulations}, we illustrate our theoretical results in comparison to simulations of the synthetic teacher-student dataset described above at finite dimension $d=1000$. We plot the cosine similarity with the teacher vector as a function of time for different values of the discrete step size $\gamma$ in the left panel and mini-batch size $b$ (fraction of samples in each mini-batch) in the right panel. We observe a perfect agreement between simulations and the theory correctly capturing the effect of the learning rate and of the mini-batch size. 

\begin{figure}[ht]
    \centering
    \includegraphics[scale=.47]{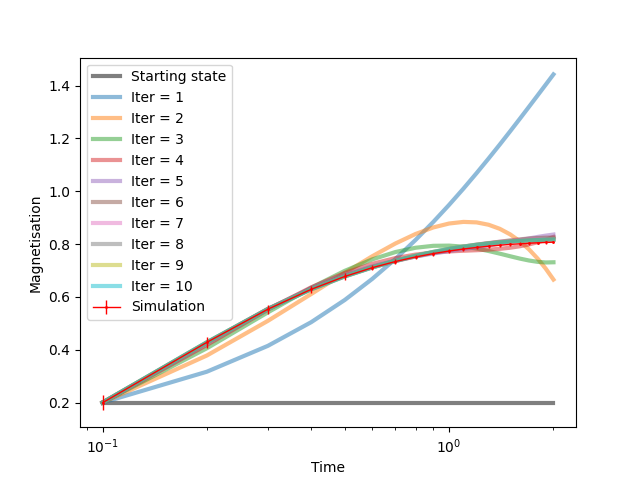}
    \caption{Evolution of the magnetization obtained from the DMFT equations as the algorithm iterates (lines). We fix the  parameters ratio of number of samples per dimension $\alpha=3$, regularization $\lambda=0.5$, the learning rate $\eta=0.1$, the mini-batch size $b=1$, the initial magnetization is $0.2$. The stochastic process in the DMFT equations is sampled more than $2500$ times for each iteration. We average the new proposal with the kernels with the previous values, keeping $70\%$ of the new kernel and $30\%$ of the old ones. 
    Points: magnetization from SGD simulations on a dataset with dimension $d=1000$. 
    }
    \label{fig:Iterations}
\end{figure}

Fig.~\ref{fig:Iterations} further illustrates the convergence of the numerical iterations solving the DMFT equations to the fixed point and agreement of this fixed point with the simulations. We observe a very fast convergence.


\section{Conclusion} 

We have proven a set of self-consistent equations characterizing the high-dimensional dynamics of first-order gradient based methods in discrete time, providing a rigorous counterpart to dynamical mean-field theory for a fairly generic family of iterations. We also provide an implementation of solver of the self-consistent equations that works well up to relatively short evolution times. 
One of the remaining key difficulties is to find stable numerical schemes to solve the DMFT equations at large times, which is a long-standing problem in this literature. Interestingly, DMFT has also been successfully applied in condensed matter physics \cite{MV89,GKKR96}, where very efficient solvers have been implemented. Also, in a wide range of realistic models, the covariance matrix of the data depends on a feature map that may change with time. Our theory currently does not allow the data matrix to have a time-dependent covariance and finding a mapping that solves this problem can be of great practical interest. Finally, from a theoretical perspective, it would be interesting to see if the DMFT equations can be simplified to extract key quantities governing the convergence of descent algorithms in high-dimension. Such an approach has been recently proposed in \cite{arous2021online,arous2022high,veiga2022phase} for online SGD, where the geometry of the landscape appears through a quantity (the information exponent) related to the higher-order derivatives of the cost function, and summary statistics of the dynamics can be chosen to study specific properties. Extending such results to full or mini-batch algorithms would be a significant step forward to better understand descent methods of practical interest.

\section*{Acknowledgments}
The authors would like to thank two anonymous referees for helpful comments on related works and technical aspects of the proof.

\newpage

\appendix

\section{Useful definitions and probability results}
\label{app:tech_app}
Here we reproduce some definitions and useful intermediate lemmas from \cite{bayati2011dynamics,gerbelot2021graph} without proof.
\begin{lemma}[Gaussian matrices under linear constraints]
    \label{lemma:cond_lemma}
    Consider an $n \times d$ random matrix $\mathbf{A}$ with i.i.d. standard normal elements, and deterministic matrices $\mathbf{Q} \in \mathbb{R}^{d \times k}$, $\mathbf{M} \in \mathbb{R}^{n \times k}$, such that the projectors 
    $\mathbf{P}_{\mathbf{M}} = \mathbf{M}\left(\mathbf{M}^{\top}\mathbf{M}\right)^{-1}\mathbf{M}^{\top}$ and $\mathbf{P}_{\mathbf{Q}} = \mathbf{Q}\left(\mathbf{Q}^{\top}\mathbf{Q}\right)^{-1}\mathbf{Q}^{\top}$ onto the subspaces spanned by the columns of 
    $\mathbf{Q}$ and $\mathbf{M}$ exist. Then the conditional distribution of $\mathbf{A}$ given the random variables $\mathbf{A}\mathbf{Q},\mathbf{A}^{\top}\mathbf{M}$ may be written
    \begin{equation}
    \mathbf{A}\vert_{\mathbf{A}\mathbf{Q},\mathbf{A}^{\top}\mathbf{M}} = \mathbf{P}_{\mathbf{M}}\mathbf{A}+\mathbf{A}\mathbf{P}_{\mathbf{Q}}-\mathbf{P}_{\mathbf{M}}\mathbf{A}\mathbf{P}_{\mathbf{Q}}+\mathbf{P}^{\perp}_{\mathbf{M}}\tilde{\mathbf{A}}\mathbf{P}^{\perp}_{\mathbf{Q}}
    \end{equation}
    where $\mathbf{P}^{\perp}_{\mathbf{M}} = \mathbf{I}_{n}-\mathbf{P}_{\mathbf{M}}$, $\mathbf{P}^{\perp}_{\mathbf{Q}} = \mathbf{I}_{d}-\mathbf{P}_{\mathbf{Q}}$, and $\tilde{\mathbf{A}}$ is an independent copy of $\mathbf{A}$.
\end{lemma}

\begin{lemma}[Gaussian concentration of pseudo-Lipschitz functions]
    \label{lemma:pseudo-lip-conv}
    Let $\bZ \sim \mathbf{N}(0, \mathbf{K}\otimes \mathbf{I}_{N})$ where $\mathbf{K} \in \cS_{q}^{+}$ has bounded operator norm. Let $\Phi_{N} : \mathbb{R}^{N\times q} \to \mathbb{R}$ be a sequence of random functions, independent of $\bZ$, such that $\mathbb{P}(\mathcal{E}_{N}) \to 1$ as $N \to \infty$, where $\mathcal{E}_{N}$ is the event that $\Phi_{N}$ is pseudo-Lipschitz of (deterministic) order $k$ with (deterministic) pseudo-Lipschitz constant $L$. Then $\Phi_{N}(\bZ) \approxP \mathbb{E}[\Phi_{N}(\bZ)]$. 
\end{lemma}

\begin{lemma}[Stein's lemma, matrix version]
    \label{matrix-stein}
    Let $(\mathbf{\bZ}_{1},\mathbf{\bZ}_{2}) \in \left(\mathbb{R}^{N \times q}\right)^{2}$ be two $\mathbf{N}(0, \mathbf{K} \otimes \mathbf{I}_{N})$ random vectors, where $\mathbf{K}\in \mathbb{R}^{(2q) \times (2q)}$
    \begin{equation}
        \mathbf{K} = \begin{bmatrix}
        \mathbf{K}_{11} \thickspace \mathbf{K}_{12} \\
        \mathbf{K}_{12} \thickspace  \mathbf{K}_{22}\end{bmatrix}
    \end{equation}
    Consider an almost everywhere differentiable function $f:\mathbb{R}^{N \times q} \to \mathbb{R}^{N \times q}$. For any $\mathbf{\bZ} \in \mathbb{R}^{N \times q}$ we can write:
    \begin{equation}
        f\left(\begin{bmatrix}
        \bZ_{11}, ..., \bZ_{1q} \\
        ...
        \\
        \bZ_{n1}, ...,\bZ_{nq}
        \end{bmatrix}\right) = \begin{bmatrix}
        f_1(\mathbf{\bZ}) \\
        ... \\
        f_{n}(\mathbf{\bZ})\end{bmatrix} = \begin{bmatrix}
        f_{1}^{1}(\mathbf{\bZ}), ...f_{1}^{q}(\mathbf{\bZ})\\
        ...\\
        f_{n}^{1}(\mathbf{\bZ}), ..., f_{n}^{q}(\mathbf{\bZ})
        \end{bmatrix}
    \end{equation}
    Then 
    \begin{equation}
        \mathbb{E}\left[(\mathbf{\bZ}_{1})^{\top}f(\mathbf{\bZ}_{2})\right] = \mathbf{K}_{1,2}\left(\sum_{k=1}^{N}\mathbb{E}\left[\frac{\partial f_{k}(\mathbf{\bZ}_{2})}{\partial \bZ_{k}}\right]\right)^{\top}
    \end{equation}
    where $\frac{\partial f_{k}(\mathbf{\bZ}_{2})}{\partial \bZ_{k}} \in \mathbb{R}^{q \times q}$ is the Jacobian containing the partial derivatives of $f_{k}$ w.r.t. the line $\mathbf{\bZ}_{k} \in \mathbb{R}^{q}$.
    \end{lemma}
\begin{lemma}[Miscellaneous results on Gaussian random matrices]
    \label{conv_lemmas_app}
    Consider a sequence of matrices $\mathbf{A} \sim GOE(N)$ and two sequences of non-random matrices, $\bU,\bV \in \mathbb{R}^{N \times q}$ such that the columns of $\bU$ and $\bV$ verify $\norm{\bU^{i}}_{2} = \norm{\bV^{i}}_{2} = \sqrt{N}$. Under this hypothesis, define the finite quantity $\mathbf{G} = \lim_{N \to \infty}\frac{1}{N}\bU^{\top}\bU$, the
    limiting Gram matrix of the columns of $\bU$. We then have: 
    \begin{enumerate}[label=\alph*)]
    \item $\frac{1}{N}\bV^{\top}\mathbf{A}\bU \xrightarrow[N \to \infty]{P} 0_{q \times q}$ and $\frac{1}{N}\norm{\bV^{\top}\mathbf{A}\bU}_{F} \xrightarrow[N \to \infty]{P} 0$.
    \item Let $\mathbf{P} \in \mathbb{R}^{N \times N}$ be a sequence of non-random projection matrices such that there exists a constant t that satisfies, for all N, k=rank($\mathbf{P}$)$\leqslant t$. Then $\frac{1}{N}\norm{ \mathbf{P}\mathbf{A}\bU}_{F}^{2} \xrightarrow[N \to \infty]{P} 0$.
    \item There exists a sequence of random matrices $\bZ \in \mathbb{R}^{N \times q}$, such that
    \begin{equation}
     \frac{1}{N} \norm{\mathbf{A}\bU-\bZ}^{2}_{F} \xrightarrow[N \to \infty]{P} 0
    \end{equation}
     where $\bZ \sim \mathbf{N}(0,\mathbf{G}\otimes \mathbf{I}_{N})$.
    \item $\frac{1}{N} (\mathbf{A}\bU)^{\top}\mathbf{A}\bU  \xrightarrow[N \to \infty]{P} \mathbf{G}$.
    \end{enumerate}
    \end{lemma}
Note that, in the proof of Theorem \ref{th:main_dmft}, we consider a random initialization matrix $\mathbf{v}^{0} \in \mathbf{R}^{d \times q}$ with i.i.d. subGaussian elements, independent from the elements of $\mathbf{G}$. The proofs of Lemma \ref{conv_lemmas_app} can be adapted straightforwardly to the case where $\mathbf{U},\mathbf{V}$ are replaced by matrices independent on $\mathbf{G}$ with i.i.d. subGaussian entries by repeating the argument conditionally on $\mathbf{U},\mathbf{V}$. The conditioning can then be lifted using concentration of inner products of subGaussian random vectors \cite{vershynin2018high}.
We include the following lemma for separable functions, which is used in the proof of Theorem \ref{th:main_dmft} under assumption $(A3.b)$ and the proof of Corollary \ref{th:main-dmft-separable}
\begin{lemma}
    \label{lemma:sep_conc}
    Let $\mathbf{Z}_{1},...,\mathbf{Z}_{q} \in \mathbb{R}^{N}$ be independent random vectors, not necessarily identically distributed, for some constant $q$ independent on $N$. Assume that, for any $1 \leq j \leq q$, the entries of $\mathbf{Z}_{j}$ are i.i.d. subGaussian.
    Then, for any pseudo-Lipschitz function of order $k$ $f:\mathbb{R}^{q} \to \mathbb{R}$, 
    \begin{equation}
        \lim_{N \to \infty}\frac{1}{N}\sum_{i=1}^{N}f(Z_{1}^{i},...,Z_{q}^{i}) \stackrel{a.s.}= \mathbb{E}\left[f(Z_{1}^{1},...,Z_{q}^{1})\right]
    \end{equation}
\end{lemma}
\begin{proof}
    Since $f$ is pseudo-Lipschitz of order $k$, there exists a constant $L$ such that, for any $X \in \mathbb{R}^{q}$
    \begin{equation}
        \abs{f(X)} \leq L(1+\norm{X}_{2}^{k}).
    \end{equation}
    Thus, for any $1 \leq i \leq N$
    \begin{align}
        \mathbb{E}\left[f(Z_{1}^{i},...,Z_{q}^{i})\right] &\leq L(1+\mathbb{E}\left[\norm{[Z_{1}^{i},...,Z_{q}^{i}]}_{2}^{k}\right]) \\
        &\leq L(1+q^{\frac{k}{2}-1}\sum_{j=1}^{q}\mathbb{E}\left[\abs{Z_{j}^{i}}^{k}\right])
    \end{align}
    where the second line is obtained by applying H\"older's inequality to the vectors $(1,...,1) \in \mathbb{R}^{q},((Z_{1}^{i})^{2},...,(Z_{q}^{i})^{2}) \in \mathbb{R}^{q}$ . Since the $Z_{j}^{i}$ are subGaussian, the quantities $\mathbb{E}\left[\abs{Z_{j}^{i}}^{k}\right]$ are 
    finite for any $1 \leq i \leq N, 1 \leq j \leq q$ and any finite $k$. We deduce that, for any $1 \leq i \leq N$, the scalar random variable 
    $G^{i} = f(Z_{1}^{i},...,Z_{q}^{i})$ has finite mean. Since the $G^{i}$ are i.i.d. with finite mean, we may conclude using Kolmogorov's strong law of large numbers.
\end{proof}
\section{Proof of Theorem \ref{th:main_dmft}}
\label{sec:app_proof}
This appendix provides the details for the second part of the induction proving Theorem \ref{th:main_dmft}, the first part of which we presented in section \ref{sec:gen_proof}. At this point we completed the induction step for the variable $\mathbf{r}^{t}$.
    Moving to $\mathbf{v}^{t+1}$, we now need to condition on $\mathfrak{S}^{t}$ but also on $\mathbf{r}^{t}$ for which we just proved the statement, which amounts to conditioning on the values of $\mathbf{v}^{0},\mathbf{X}^{\top}\mathbf{m}^{0}, ...,\mathbf{X}^{\top}\mathbf{m}^{t-1}$, $\mathbf{X}\mathbf{w}^{0}, ..., \mathbf{X}\mathbf{w}^{t}$. We denote 
     $\tilde{\mathfrak{S}}_{t}$ the smallest $\sigma-$ algebra containing $\mathfrak{S}_{t}$ and $\sigma(\mathbf{r}^{t})$, the $\sigma$-algebra generated by $\mathbf{r}^{t}$.
    We will then perform orthogonal decompositions on the subspaces spanned by the matrices
    \begin{equation}
    \mathbf{M}_{t-1} = \left[\mathbf{m}^{0} \vert \mathbf{m}^{1} \vert ... \vert \mathbf{m}^{t-1} \right], \mathbf{W}_{t} = \left[\mathbf{w}^{0} \vert \mathbf{w}^{1} \vert ... \vert \mathbf{w}^{t-1} \vert \mathbf{w}^{t}\right]
    \end{equation}
    where $\mathbf{M}_{t-1} \in \mathbb{R}^{n \times tq}$ and $\mathbf{W}_{t} \in \mathbb{R}^{d \times tq}$. Using lemma \ref{lemma:cond_lemma} and the fact that $\bm{h}^{t}(\left\{\mathbf{v}^{k}\right\}_{k=0}^{t}), \mathbf{m}^{t}$ are $\tilde{\mathfrak{S}}^{t}-$measurable, we obtain
    \begin{align}
        &\mathbf{v}^{t+1}\vert_{\tilde{\mathfrak{S}}^{t}} \stackrel{d} = \bm{h}^{t}(\left\{\mathbf{v}^{k}\right\}_{k=0}^{t})+\mathbf{X}\vert_{\tilde{\mathfrak{S}}^{t}}^{\top}\mathbf{m}^{t} \\
        & \stackrel{d}= \bm{h}^{t}(\left\{\mathbf{v}^{k}\right\}_{k=0}^{t})+\left(\mathbf{X}^{\top}\mathbf{P}_{\mathbf{M}_{t-1}}+\mathbf{P}_{\mathbf{W}_{t}}\mathbf{X}^{\top}-\mathbf{P}_{\mathbf{W}_{t}}\mathbf{X}^{\top}\mathbf{P}_{\mathbf{M}_{t-1}}+\mathbf{P}^{\perp}_{\mathbf{W}_{t}}\tilde{\mathbf{X}}^{\top}\mathbf{P}^{\perp}_{\mathbf{M}_{t-1}}\right)\mathbf{m}^{t} \\
        &= \bm{h}^{t}(\left\{\mathbf{v}^{k}\right\}_{k=0}^{t})+\mathbf{X}^{\top}\mathbf{P}_{\mathbf{M}_{t-1}}\mathbf{m}^{t}+\mathbf{P}_{\mathbf{W}_{t}}\mathbf{X}^{\top}\mathbf{P}_{\mathbf{M}_{t-1}}^{\perp}\mathbf{m}^{t}+\mathbf{P}^{\perp}_{\mathbf{W}_{t}}\tilde{\mathbf{X}}^{\top}\mathbf{P}^{\perp}_{\mathbf{M}_{t-1}}\mathbf{m}^{t} \label{eq:interb2}
    \end{align}
    where $\tilde{\mathbf{X}}$ is an independent copy of $\mathbf{X}$.
    As before, we treat each term separately, starting with $\bm{h}^{t}(\left\{\mathbf{v}^{k}\right\}_{k=0}^{t})$, for which the induction hypothesis gives 
    \begin{equation}
        \label{eq:b1_1}
        \frac{1}{\sqrt{d}}\norm{\bm{h}^{t}(\left\{\mathbf{v}^{k}\right\}_{k=0}^{t})-\mathbf{h}^{t}(\{\bnu^{k}\}_{k=0}^{t})}_{F} \xrightarrow[n,d \to \infty]{P} 0,
    \end{equation}
where the $\{\bnu^{k}\}_{k=0}^{t}$ are defined as in Theorem \ref{th:main_dmft}. Moving to the second term in Eq.\eqref{eq:interb2},
    \begin{align}
        &\mathbf{X}^{\top}\mathbf{P}_{\mathbf{M}_{t-1}}\mathbf{m}^{t}=\mathbf{X}^{\top}\mathbf{M}_{t-1}\left(\mathbf{M}_{t-1}^{\top}\mathbf{M}_{t-1}\right)^{-1}\mathbf{M}_{t-1}^{\top}\mathbf{m}^{t} \\
        &=\left[\mathbf{v}^{1}-\bm{h}^{0}(\mathbf{w}^{0})\vert ... \vert \mathbf{v}^{t}-\bm{h}^{t-1}(\left\{\mathbf{v}^{k}\right\}_{k=0}^{t-1}) \right]^{\top}\boldsymbol{\beta}^{t} \\
        &= \sum_{k=0}^{t-1}\left(\mathbf{v}^{k+1}-\bm{h}^{t}(\left\{\mathbf{v}^{l}\right\}_{l=0}^{k})\right)\boldsymbol{\beta}^{t}_{k}
    \end{align}
    where, for any we defined the $tq\times q$ matrix containing the projection coefficients of $\mathbf{m}^{t}$ on the subspace spanned by the columns of $\mathbf{M}_{t-1}$:
    \begin{align}
    \boldsymbol{\beta}^{t} &= \left(\mathbf{M}_{t-1}^{\top}\mathbf{M}_{t-1}\right)^{-1}\mathbf{M}_{t-1}^{\top}\mathbf{m}^{t}
    \end{align}
    and, for any $0 \leqslant k \leqslant t$, $\beta_{k}^{t}$ denotes the $k-th$ block of size $q \times q$ of $\boldsymbol{\beta}^{t}$.
    Using the induction hypothesis and the non-degeneracy assumption, we have the following convergence result for $\boldsymbol{\beta}^{t}$
    \begin{align}
    \boldsymbol{\beta}^{t}&= \left(\frac{1}{n}\mathbf{M}_{t-1}^{\top}\mathbf{M}_{t-1}\right)^{-1}\frac{1}{n}\mathbf{M}_{t-1}^{\top}\mathbf{m}^{t}\\
    &\approxP \boldsymbol{\beta}^{t,*} \in \mathbb{R}^{tq \times q}
    \end{align}
    with deterministic $\boldsymbol{\beta}^{t,*}$, in similar fashion to the claim for $\boldsymbol{\alpha}^{t,*}$. Letting \\ $\mathbf{G}_{t-1} = [\mathbf{g}^{0}(\beeta^{0}) \vert ... \vert \mathbf{g}^{t-1}(\beeta^{t-1})]$, we also have the following expression for $\boldsymbol{\beta}^{t,*}$:
    \begin{equation}
        \boldsymbol{\beta}^{t,*} \approxP (\mathbf{G}_{t-1}^{\top}\mathbf{G}_{t-1})^{-1}(\mathbf{G}_{t-1})^{\top}\mathbf{g}^{t}(\beeta^{t}).
    \end{equation}
    A straightforward application of the triangle inequality along with the induction hypothesis then leads to
    \begin{align}
        \label{eq:b1_2}
        &\frac{1}{\sqrt{n}}\norm{\mathbf{X}^{\top}\mathbf{P}_{\mathbf{M}_{t-1}}\mathbf{m}^{t}-\sum_{k=0}^{t-1}\left(\btheta^{k}\Gamma^{k}+\sum_{l=0}^{k-1}\btheta^{l}R_{g}(k,l)+\mathbf{u}^{k}\right)\beta^{*,t}_{k}}_{F} \xrightarrow[n,d \to \infty]{P} 0.
    \end{align}
Moving to the third term in Eq.\eqref{eq:inter2}, we write
    \begin{align}
        &\mathbf{P}_{\mathbf{W}_{t}}\mathbf{X}^{\top}\mathbf{P}_{\mathbf{M}_{t-1}}^{\perp}\mathbf{m}^{t} = \mathbf{W}_{t-1}\left(\mathbf{W}_{t}^{\top}\mathbf{W}_{t}\right)^{-1}\mathbf{W}_{t}^{\top}\mathbf{X}^{\top}\mathbf{P}_{\mathbf{M}_{t-1}}^{\perp}\mathbf{m}^{t} \\
        &=\mathbf{W}_{t}\left(\mathbf{W}_{t}^{\top}\mathbf{W}_{t}\right)^{-1}\left[\mathbf{r}^{0}\vert ... \vert \mathbf{r}^{t}\right]^{\top}\mathbf{P}_{\mathbf{M}_{t-1}}^{\perp}\mathbf{m}^{t}.
    \end{align}
    Using a similar argument as in the proof of the induction step for $\mathbf{r}^{t}$, we may use the induction hypothesis and non-degeneracy assumption to write the limiting 
    behaviour of the projector $\mathbf{P}_{\mathbf{M}_{t-1}}^{\perp}$ to obtain
    \begin{align}
        &\frac{1}{n}\left[\mathbf{r}^{0}\vert ... \vert \mathbf{r}^{t-1}\right]^{\top}\mathbf{P}_{\mathbf{M}_{t-1}}^{\perp}\mathbf{m}^{t} \approxP \frac{1}{d}\left[\bomega^{0} \vert ... \vert \bomega^{t}\right]^{\top}\mathbf{P}_{\mathbf{G}_{t-1}}^{\perp}\mathbf{g}^{t}(\beeta^{t}) \\
        &= \frac{1}{n}\left[\bomega^{0} \vert ... \vert \bomega^{t}\right]^{\top}\mathbf{g}^{t}(\beeta^{t})-\frac{1}{d}\left[\bomega^{0} \vert ... \vert \bomega^{t}\right]^{\top}\mathbf{P}_{\mathbf{G}_{t-1}}\mathbf{g}^{t}(\beeta^{t}) \\
        &\approxP \frac{1}{n}\mathbb{E}\left[\left[\bomega^{0} \vert ... \vert \bomega^{t}\right]^{\top}\mathbf{g}^{t}(\beeta^{t})\right]-\frac{1}{n}\mathbb{E}\left[\left[\bomega^{0} \vert ... \vert \bomega^{t}\right]^{\top}\left[\mathbf{g}^{0}(\beeta^{0}) \vert ... \vert \mathbf{g}^{t-1}(\beeta^{t-1})\right]\right]\boldsymbol{\beta}^{t,*},
    \end{align}
    where, for any $0 \leqslant s \leqslant t$, Stein's lemma gives
    \begin{align}
        \frac{1}{n}\mathbb{E}\left[\left(\bomega^{s}\right)^{\top}\mathbf{m}^{t}\right] = \frac{1}{n}\mathbb{E}\left[\left(\bomega^{s}\right)^{\top}\bm{g}^{t}\left(\beeta^{t}\left(\bomega^{0},..., \bomega^{t-1},\bomega^{t}\right)\right)\right] = \frac{1}{n}\sum_{i=0}^{t}C_{\theta}(s,i)\sum_{j=1}^{n}\mathbb{E}\left[\frac{\partial g^{t}_{j}}{\partial \omega^{i}_{j}}(\eta^{t})\right],
    \end{align}
    and where, for any $0 \leqslant i \leqslant t$ and $0 \leqslant j \leqslant n$, $\frac{\partial g^{t}_{j}}{\partial \omega^{i}_{j}}(\eta^{t})$ denotes the $q \times q$ jacobian matrix containing the partial derivatives of the restriction of $\mathbf{g}^{t}(\boldsymbol{\eta}^{t}(.))$ to the $j-$th line of its output, with respect to the $j-$th line of $\bomega^{i}$.
From the definition of $\beeta^{t}$ in Theorem \ref{th:main_dmft}, the dependence on $\bomega^{t}$ in $\beeta^{t}$ is the identity. We may then write 
  \begin{align}
        \frac{1}{n}\mathbb{E}\left[\frac{\partial g^{t}_{j}}{\partial \omega^{t}_{j}}(\eta^{t})\right] =  \frac{1}{n}\sum_{j=1}^{n}\mathbb{E}\left[\frac{d g^{t}_{j}}{d \eta^{t}_{j}}(\eta^{t})\right] = \Gamma^{t}
    \end{align}
We now define $\mathbf{C}_{\theta,t}$ as the $(t+1)q\times (t+1)q$ covariance matrix of the lines of $\left[\bomega^{0}\vert... \vert \bomega^{t-1} \vert \bomega^{t}\right] \in \mathbb{R}^{n \times (t+1)q}$, and 
\begin{equation}
\mathbf{R}_{g,t} = \begin{bmatrix}\frac{1}{n}\sum_{j=1}^{n}\mathbb{E}\left[\frac{\partial g^{t}_{j}}{\partial \bomega^{0}_{j}}(\eta^{t})\right] \\
... \\
\frac{1}{n}\sum_{j=1}^{n}\mathbb{E}\left[\frac{\partial g^{t}_{j}}{\partial \omega^{t-1}_{j}}(\beeta^{t})\right] \\
\frac{1}{n}\sum_{j=1}^{n}\mathbb{E}\left[\frac{d g^{t}_{j}}{d \beeta^{t}_{j}}(\beeta^{t})\right]
\end{bmatrix} \in \mathbb{R}^{(t+1)q \times q}.
\end{equation}
We thus have
\begin{equation}
\frac{1}{n}\mathbb{E}\left[\left[\bomega^{0} \vert ... \vert \bomega^{t-1}\right]^{\top}\mathbf{m}^{t}\right] = \mathbf{C}_{\theta,t}\mathbf{R}_{g,t},
\end{equation}
and, for any $0\leqslant s < t$
\begin{equation}
\frac{1}{n}\mathbb{E}\left[\left[\bomega^{0} \vert ... \vert \bomega^{t-1}\right]^{\top}\mathbf{m}^{s}\right] = \mathbf{C}_{\theta,t}\begin{bmatrix}\frac{1}{n}\sum_{j=1}^{n}\mathbb{E}\left[\frac{\partial g^{s}_{j}}{\partial \omega^{0}_{j}}(\eta^{s})\right] \\
... \\
\frac{1}{n}\sum_{j=1}^{n}\mathbb{E}\left[\frac{\partial g^{s}_{j}}{\partial \omega^{s-1}_{j}}(\eta^{s})\right]  \\
\frac{1}{n}\sum_{j=1}^{n}\mathbb{E}\left[\frac{d g^{s}_{j}}{d \eta^{s}_{j}}(\eta^{s})\right]  \\
0 \\
... \\
0
\end{bmatrix} = \mathbf{C}_{\theta,t}\mathbf{R}_{g,s}
\end{equation}
where the zeroes come from the fact that $\beeta^{s}$ is not an algebraic function of the $\bomega^{l}$ for $l > s$ which is, again, coherent with notions of causality. Note that the matrices $\mathbf{R}_{g,s}$ are defined in such a way that, for any $0 \leqslant s \leqslant t$, $\mathbf{R}_{g,s}$ all have the same dimension $tq \times q$. We thus reach the following equality
\begin{align}
    &\frac{1}{n}\mathbb{E}\left[\left[\bomega^{0} \vert ... \vert \bomega^{t}\right]^{\top}\mathbf{m}^{t}\right]-\frac{1}{n}\mathbb{E}\left[\left[\bomega^{0} \vert ... \vert \bomega^{t}\right]^{\top}\mathbf{M}_{t-1}\right]\boldsymbol{\beta}^{t,*} \\
    &=\mathbf{C}_{\theta,t}\left(\mathbf{R}_{g,t}-\left[\mathbf{R}_{g,0}\vert \mathbf{R}_{g,1} \vert ... \vert \mathbf{R}_{g,t-1}\right]\boldsymbol{\beta}^{t,*}\right).
\end{align}
Also, due to the induction hypothesis 
\begin{equation}
\frac{1}{n}\mathbf{W}^{T}_{t}\mathbf{W}_{t} \approxP \mathbf{C}_{\theta,t}
\end{equation}
which leads to 
\begin{align}
    \left(\mathbf{W}^{T}_{t}\mathbf{W}_{t}\right)^{-1}\left[\mathbf{r}^{0}\vert ... \vert \mathbf{r}^{t-1}\right]^{\top}\mathbf{P}_{\mathbf{M}_{t-1}}^{\perp}\mathbf{m}^{t} \approxP \mathbf{R}_{g,t}-\left[\mathbf{R}_{g,0}\vert \mathbf{R}_{g,1} \vert ... \vert \mathbf{R}_{g,t-1}\right]\boldsymbol{\beta}^{t,*}
\end{align}
and
\begin{align}
\frac{1}{\sqrt{d}}\norm{\mathbf{P}_{\mathbf{W}_{t}}\mathbf{X}^{\top}\mathbf{P}_{\mathbf{M}_{t-1}}^{\perp}\mathbf{m}^{t}-\mathbf{W}_{t}\left(\mathbf{R}_{g,t}-\left[\mathbf{R}_{g,0}\vert \mathbf{R}_{g,1} \vert ... \vert \mathbf{R}_{g,t-1}\right]\boldsymbol{\beta}^{t,*}\right)}_{F} \xrightarrow[n,d \to \infty]{P} 0.
\end{align}
We may now use the induction hypothesis to obtain 
\begin{equation}
    \label{eq:b1_3}
    \frac{1}{\sqrt{d}}\norm{\mathbf{W}_{t}\left[\mathbf{R}_{g,0}\vert \mathbf{R}_{g,1} \vert ... \vert \mathbf{R}_{g,t-1}\right]\boldsymbol{\beta}^{t,*}-\sum_{k=0}^{t-1}\left(\btheta^{k}\Gamma^{k}+\sum_{l=0}^{k-1}\btheta^{l}R_{g}(k,l)\right)\beta^{*,t}_{k}}_{F} \xrightarrow[n,d \to \infty]{P} 0
\end{equation}
and 
\begin{equation}
    \label{eq:b1_3bis}
    \frac{1}{\sqrt{d}}\norm{\mathbf{W}_{t}\mathbf{R}_{g,t}-\sum_{k=0}^{t-1}\btheta^{k}R_{g}(t,k)-\btheta^{t}\Gamma^{t}}_{F} \xrightarrow[n,d \to \infty]{P} 0
\end{equation}
    where we remind that, for any $s<t$, the elements of the last $q \times q$ block of $\mathbf{R}_{g,s}$ are all zeroes, and thus $\mathbf{w}^{t}$ does not appear in the corresponding sums.
    Finally, we turn to the fourth term in Eq.\eqref{eq:interb2}. Using the fact that  $\tilde{\mathbf{X}}$ is independent of $\tilde{\mathfrak{S}}_{t}$, we may use lemma \ref{conv_lemmas_app} to show that 
    \begin{equation}
        \frac{1}{\sqrt{d}}\norm{\mathbf{P}^{\perp}_{\mathbf{W}_{t-1}}\tilde{\mathbf{X}}^{\top}\mathbf{P}^{\perp}_{\mathbf{M}_{t-1}}\mathbf{m}^{t}-\tilde{\mathbf{X}}^{\top}\mathbf{P}^{\perp}_{\mathbf{M}_{t-1}}\mathbf{m}^{t}}_{F} \xrightarrow[n,d \to \infty]{P} 0,
    \end{equation}
    and use the induction hypothesis to show that there exists a $d \times q$ random matrix $\tilde{\mathbf{u}}^{t}$ distributed according to $\mathcal{N}(0,C^{\perp}_{\mathbf{m},t} \otimes \mathbf{I}_{d})$, such that 
    \begin{equation}
        \label{eq:b1_4}
        \frac{1}{\sqrt{d}}\norm{\tilde{\mathbf{X}}^{\top}\mathbf{P}^{\perp}_{\mathbf{M}_{t-1}}\mathbf{m}^{t}-\tilde{\mathbf{u}}^{t}}_{F} \xrightarrow[n,d \to \infty]{P} 0
    \end{equation}
    where
    \begin{equation}
        C^{\perp}_{\mathbf{m},t} = \lim_{n,d \to \infty} \frac{1}{n}\left(\mathbf{P}^{\perp}_{\mathbf{M}_{t-1}}\mathbf{m}^{t}\right)^{\top}\mathbf{P}^{\perp}_{\mathbf{M}_{t-1}}\mathbf{m}^{t}
        \end{equation}
        and $\tilde{u}^{t}$ is independent from all other random parameters of the problem. 
        Combining Eq.\eqref{eq:b1_1},\eqref{eq:b1_2}, \\\eqref{eq:b1_3},\eqref{eq:b1_3bis} and \eqref{eq:b1_4} with the triangle inequality, we reach the following asymptotic representation of $\mathbf{v}^{t+1}\vert_{\tilde{\mathfrak{S}}^{t}}$
        \begin{align}
        \label{eq:conv2}
            \frac{1}{\sqrt{d}}\norm{\mathbf{v}^{t+1}\vert_{\tilde{\mathfrak{S}}^{t}}-\left(\bm{h}^{t}(\boldsymbol{\omega}^{t})+\btheta^{t}\Gamma^{t}+\sum_{k=0}^{t-1}\btheta^{k}R_{g}(t,k)+\sum_{k=0}^{t-1}\mathbf{u}^{k}\beta^{*,t}_{k}+\tilde{\mathbf{u}}^{t}\right)}_{F} \xrightarrow[n,d \to \infty]{P} 0,
        \end{align}
        which matches the equation for the asymptotic representation of $\mathbf{v}^{t+1}$ from Theorem \ref{th:main_dmft}, provided the Gaussian process term has the correct covariance. The statement that, for any sequence of pseudo-Lipschitz functions $\{\phi_{n}\}_{n >0}$
        \begin{align}
            \phi_{n}(\mathbf{v}_{0},\mathbf{v}^{1},...,\mathbf{v}^{t+1})\vert_{\tilde{\mathfrak{S}}^{t}} \approxP \phi_{n}(\mathbf{v}^{0},\mathbf{v}^{1},...,\bm{h}^{t}(\boldsymbol{\omega}^{t})+\btheta^{t}\Gamma^{t}+\sum_{k=0}^{t-1}\btheta^{k}R_{g}(t,k)+\sum_{k=0}^{t-1}\mathbf{u}^{k}\beta^{*,t}_{k}+\tilde{\mathbf{u}}^{t}),
        \end{align}
        is proven in similar fashion to the corresponding step in the induction step on $\mathbf{r}^{t}$ using the induction hypothesis, definition of pseudo-Lipschitz function and Eq.\eqref{eq:conv2}.
We now turn to verifying the covariance profile of the additive Gaussian process term $\sum_{k=0}^{t-1}\mathbf{u}^{k}\beta^{*,t}_{k}+\tilde{\mathbf{u}}^{t}$. Define 
\begin{equation}
\mathbf{u}^{t} = \sum_{k=0}^{t-1}\mathbf{u}^{k}\beta^{*,t}_{k}+\tilde{\mathbf{u}}^{t}
\end{equation}
To check the $\mathbf{u}^{t}$ has the correct covariance profile, we evaluate, for any $s<t$
\begin{align}
\frac{1}{d}\mathbb{E}\left[(\mathbf{u}^{s})^{\top}\mathbf{u}^{t}\right] &= \sum_{k=0}^{t-1}\mathbb{E}\left[(\mathbf{u}^{s})^{\top}\mathbf{u}^{k}\right]\beta^{*,t}_{k} \\
&=\mathbf{C}_{g,t}\boldsymbol{\beta}^{*,t} \\
&\approxP \frac{1}{d}(\mathbf{m}^{s})^{\top}\mathbf{M}_{t-1}\left(\mathbf{M}_{t-1}^{\top}\mathbf{M}_{t-1}\right)^{-1}\mathbf{M}_{t-1}^{\top}\mathbf{m}^{t} \\
&\approxP\frac{1}{d}(\mathbf{m}^{s})^{\top}\mathbf{m}^{t} \\
&\approxP \frac{1}{d}\mathbb{E}\left[\mathbf{g}^{s}(\beeta^{s})^{\top}\mathbf{g}^{t}(\beeta^{t})\right]
\end{align}
and for $s=t$
\begin{align}
\frac{1}{d}\mathbb{E}\left[(\mathbf{u}^{t})^{\top}\mathbf{u}^{t}\right] &= \sum_{k=0}^{t-1}\sum_{k'=0}^{t-1}(\beta^{*,t}_{k})^{\top}\frac{1}{d}\mathbb{E}\left[(\mathbf{u}_{k})^{\top}\mathbf{u}_{k'}\right]\beta^{*,t}_{k'}+\frac{1}{d}\mathbb{E}\left[(\tilde{\mathbf{u}}^{t})^{\top}\tilde{\mathbf{u}}^{t}\right] \\
&\approxP \frac{1}{d}(\mathbf{m}^{t})^{\top}\mathbf{M}_{t-1}\left(\mathbf{M}_{t-1}^{\top}\mathbf{M}_{t-1}\right)^{-1}\mathbf{M}_{t-1}^{\top}\mathbf{m}^{t}+\frac{1}{d}\mathbf{m}^{t}\mathbf{P}_{\mathbf{M}_{t-1}}^{\perp}\mathbf{m}^{t} \\
&\approxP \frac{1}{d}(\mathbf{m}^{t})^{\top}\mathbf{m}^{t} \\
&\approxP \frac{1}{d}\mathbb{E}\left[\mathbf{g}^{t}(\beeta^{t})^{\top}\mathbf{g}^{t}(\beeta^{t})\right]
\end{align}
which concludes the induction.
\subsection{Relaxing the non-degeneracy assumption}
\label{app_sub:relax}
The non-degeneracy assumption can be relaxed using the same method as in \cite{berthier2020state,gerbelot2021graph}. We can define an auxiliary, randomly perturbed iteration with 
\begin{align}
        \label{eq:the_dynamics1_pert}
        \hat{\mathbf{v}}^{t+1} &= \hat{\mathbf{h}}^{t}\left(\left\{\hat{\mathbf{v}}^{k}\right\}_{k=0}^{t}\right)+\mathbf{X}^{\top}\hat{\mathbf{g}}^{t}(\hat{\mathbf{r}}^{t})  \\
        \hat{\mathbf{r}}^{t} &= \mathbf{X}\sum_{k=0}^{t}\hat{\mathbf{v}}^{k}
        \label{eq:the_dynamics2_pert}
    \end{align}
    initialized with the same $\mathbf{v}_{0}$ as the original dynamics Eq.~\eqref{eq:the_dynamics1}-\eqref{eq:the_dynamics2}, and where the update functions are defined as 
    \begin{align}
    \hat{\mathbf{h}}^{t}\left(\left\{\hat{\mathbf{v}}^{k}\right\}_{k=0}^{t}\right) &= \mathbf{h}^{t}\left(\left\{\hat{\mathbf{v}}^{k}\right\}_{k=0}^{t}\right)+\epsilon\mathbf{Y}_{h}^{t} \\
    \hat{\mathbf{g}}^{t}(\hat{\mathbf{r}}^{t}) &= \mathbf{g}^{t}(\hat{\mathbf{r}}^{t})+\epsilon\mathbf{Y}_{r}^{t}
    \end{align}
    where, at each time step, $\mathbf{Y}_{h}^{t} \in \mathbb{R}^{d \times q}$ and $\mathbf{Y}_{r}^{t} \in \mathbb{R}^{n \times q}$ have i.i.d. standard normal elements and are independent from one another and from all other parameters from the problems. Since $n,d$ are much larger than $tq$ by assumption, standard results on Gaussian matrices \cite{vershynin2018high} show that the Gram matrices being inverted in the projectors are almost surely full rank with smallest eigenvalue bounded away from $0$ when $n,d$ go to infinity. We thus have the rigorous system of equations for the perturbed iteration. Using another induction, one can then show that the iterates of the perturbed iterations uniformly converge to the original ones when taking $\epsilon$ to zero. Similarly, uniform convergence of the asymptotic Gaussian model of the perturbed iteration towards the one of the original iteration can be shown. Taking the limits on both sides concludes the proof. Since the procedure and technical steps are almost identical to those presented in \cite{berthier2020state,gerbelot2021graph}, we do not reproduce them here.

    \subsection{Proof of Corollary \ref{th:main-dmft-separable}}
    \label{app:proof_coroll}
To prove Corollary $\ref{th:main-dmft-separable}$, we show that, under assumptions (B1)-(B4), the iteration defined by Eq.\eqref{eq:sgd_iteration} is covered by Theorem \ref{th:main_dmft}.
Assumptions (A1),(A2) and (A4) are immediately verified under (B1)-(B4). We are left with checking assumption (A3.b). Since $\mathcal{L},\mathbf{F}$ are separable, so are their gradients, 
whose components are given by $l',f'$, which are pseudo-Lipschitz of order $k$ by assumption. We may then identify 
$h^{t}_{i}(w^{t}) = -\gamma f'(w^{t})$ and $g^{t}_{i}(w^{t}) = -\gamma s_{i}^{t}l'(w^{t})$, where $g^{t}_{i}$ is a pseudo-Lipschitz function of $s_{i}^{t},r^{t}$. Finally, 
by definition of the iteration Eq.\eqref{eq:sgd_iteration}, the vector $\mathbf{s}^{t}$ has i.i.d. subGaussian (actually i.i.d. bounded) entries, resampled independently at each time step.
\section{Detailed mapping for Nesterov acceleration}
\label{sec:Nesterov_detail}
Recall the equations for Nesterov accelerated gradient 
\begin{align}
    \mathbf{y}^{t} &= \mathbf{w}^{t}+\tau^{t}(\mathbf{z}^{t}-\mathbf{w}^{t}) \\
    \mathbf{w}^{t+1} &=\mathbf{y}^{t}-\gamma^{t}\left(\XX^{\top}\nabla \mathcal{L}(\XX\yy^{t})+\nabla \reg(\yy^{t})\right) \\
    \mathbf{z}^{t+1} &= \mathbf{z}^{t}+\mu^{t}\left(\mathbf{y}^{t}-\mathbf{z}^{t}\right)-\alpha^{t}\left(\XX^{\top}\nabla \mathcal{L}(\XX\yy^{t})+\nabla \reg(\yy^{t})\right)\label{eq:nesterov2}
\end{align}
Replacing $\mathbf{y}^{t}$ using its definition leads to 
\begin{align*}
    \mathbf{w}^{t+1} &=\mathbf{w}^{t}+\tau^{t}(\mathbf{z}^{t}-\mathbf{w}^{t})-\gamma^{t}\left(\XX^{\top}\nabla \mathcal{L}(\XX\left(\mathbf{w}^{t}+\tau^{t}(\mathbf{z}^{t}-\mathbf{w}^{t})\right))+\nabla \reg(\mathbf{w}^{t}+\tau^{t}(\mathbf{z}^{t}-\mathbf{w}^{t}))\right) \\
    \mathbf{z}^{t+1} &= \mathbf{z}^{t}+\mu^{t}\left(\mathbf{w}^{t}+\tau^{t}(\mathbf{z}^{t}-\mathbf{w}^{t})-\mathbf{z}^{t}\right)\\
    &\hspace{3cm}-\alpha^{t}\left(\XX^{\top}\nabla \mathcal{L}\left(\XX\left(\mathbf{w}^{t}+\tau^{t}(\mathbf{z}^{t}-\mathbf{w}^{t})\right)\right)+\nabla \reg\left(\mathbf{w}^{t}+\tau^{t}(\mathbf{z}^{t}-\mathbf{w}^{t})\right)\right)
\end{align*}
Define the variables $ \mathbf{u}^{t+1} = \mathbf{w}^{t+1}-\mathbf{w}^{t} \in \mathbb{R}^{d}, \tilde{\mathbf{u}}^{t+1} = \mathbf{z}^{t+1}-\mathbf{z}^{t} \in \mathbb{R}^{d},\mathbf{v}^{t} = \left[\mathbf{u}^{t} \vert \tilde{\mathbf{u}}^{t}\right] \in \mathbb{R}^{d \times 2},\mathbf{x}^{t} = \left[\mathbf{w}^{t} \vert \mathbf{z}^{t}\right] = \sum_{k=0}^{t}\mathbf{v}^{k} \in \mathbb{R}^{d \times 2}$. Using these variables, we may write 
\begin{align*}
    \tau^{t}(\mathbf{z}^{t}-\mathbf{w}^{t}) = \sum_{k=0}^{t}\mathbf{v}^{k}\begin{bmatrix}-\tau^{t} \\
    \tau^{t}\end{bmatrix} \\
    \mathbf{X}\left(\mathbf{w}^{t}+\tau^{t}(\mathbf{z}^{t}-\mathbf{w}^{t})\right) = \left(\mathbf{X}\sum_{k=0}^{t}\mathbf{v}^{k}\right)\begin{bmatrix}1-\tau^{t} \\
    \tau^{t}\end{bmatrix} \\
   \mu^{t}(\mathbf{w}^{t}+\tau^{t}(\mathbf{z}^{t}-\mathbf{w}^{t})-\mathbf{z}^{t}) =  \sum_{k=0}^{t}\mathbf{v}^{k}\begin{bmatrix}\mu^{t}(1-\tau^{t}) \\ \mu^{t}(\tau^{t}-1) \end{bmatrix}
\end{align*}
Defining $\mathbf{r}^{t} = \mathbf{X}\sum_{k=0}^{t}\mathbf{v}^{k}$, we obtain
\begin{align}
    \mathbf{v}^{t+1} &= \left[\sum_{k=0}^{t}\mathbf{v}^{k}\begin{bmatrix}-\tau^{t} \\
    \tau^{t}\end{bmatrix} \vert \sum_{k=0}^{t}\mathbf{v}^{k}\begin{bmatrix}\mu^{t}(1-\tau^{t}) \\ \mu^{t}(\tau^{t}-1) \end{bmatrix}\right]  \\
    &\hspace{1cm}+\left[-\gamma^{t}\nabla F\left(\sum_{k=0}^{t}\mathbf{v}^{k}\begin{bmatrix}1-\tau^{t} \\
    \tau^{t}\end{bmatrix}\right) \vert -\alpha^{t}\nabla F\left(\sum_{k=0}^{t}\mathbf{v}^{k}\begin{bmatrix}1-\tau^{t} \\
    \tau^{t}\end{bmatrix}\right) \right]  \\
    &\hspace{1cm}+\mathbf{X}^{\top}\left[-\gamma^{t}\nabla \mathcal{L}\left(\mathbf{r}^{t}\begin{bmatrix}1-\tau^{t} \\
    \tau^{t}\end{bmatrix}\right) \vert -\alpha^{t}\nabla \mathcal{L}\left(\mathbf{r}^{t}\begin{bmatrix}1-\tau^{t} \\
    \tau^{t}\end{bmatrix}\right)\right] \\
    \mathbf{r}^{t} &= \mathbf{X}\sum_{k=0}^{t}\mathbf{v}^{k}
\end{align}
which fits the form of Eq.~(\ref{eq:the_dynamics1}-\ref{eq:the_dynamics2}) by defining 
\begin{align}
    &\mathbf{h}^{t}:\mathbb{R}^{d \times 2(t+1)} \to \mathbb{R}^{d\times 2} \\
    &\left\{\mathbf{v}^{k}\right\}_{k=0}^{t} \to \left[\sum_{k=0}^{t}\mathbf{v}^{k}\begin{bmatrix}-\tau^{t} \\
    \tau^{t}\end{bmatrix} \vert \sum_{k=0}^{t}\mathbf{v}^{k}\begin{bmatrix}\mu^{t}(1-\tau^{t}) \\ \mu^{t}(\tau^{t}-1) \end{bmatrix}\right]  \\
    &\hspace{1cm}+\left[-\gamma^{t}\nabla F\left(\sum_{k=0}^{t}\mathbf{v}^{k}\begin{bmatrix}1-\tau^{t} \\
    \tau^{t}\end{bmatrix}\right) \vert -\alpha^{t}\nabla F\left(\sum_{k=0}^{t}\mathbf{v}^{k}\begin{bmatrix}1-\tau^{t} \\
    \tau^{t}\end{bmatrix}\right) \right]  \\
    &\mathbf{g}^{t} : \mathbb{R}^{n \times 2} \to \mathbb{R}^{n \times 2} \\
    &\mathbf{r}^{t} \to \left[-\gamma^{t}\nabla \mathcal{L}\left(\mathbf{r}^{t}\begin{bmatrix}1-\tau^{t} \\
    \tau^{t}\end{bmatrix}\right) \vert -\alpha^{t}\nabla \mathcal{L}\left(\mathbf{r}^{t}\begin{bmatrix}1-\tau^{t} \\
    \tau^{t}\end{bmatrix}\right)\right]
\end{align}
\section{Details on the numerics} 
\label{app:numerics}
In this appendix, we provide additional details on the numerical solution of the DMFT equations. We start by presenting an efficient 
simplification of the system of equations in corollary \ref{th:main-dmft-separable}, that allows to reduce the number of kernels and auxiliary functions that must be computed self-consistently in the numerics. This is the 
system of the equations that we implement in the code available at \url{https://github.com/SPOC-group/Rigorous-dynamical-mean-field-theory}. We focus on the teacher-student perceptron setting introduced in section \ref{sec:numerics} and on a multi-pass SGD dynamics with ridge regularisation of strength $\lambda\geq 0$. We derive a closed system of equations for the effective low-dimensional description of a coordinate of the pre-activation term $\bm{r}^t=\bm{X}\bm{w}^t$, i.e., the relevant variable capturing the learning properties. Consider the dynamics in Eq.~\eqref{eq:sgd_iteration} projected on the direction of a training sample $\bm{x}_\mu\sim\mathcal{N}(\bm{0},\frac 1d \bm{I})$, $\mu\in\{1,\ldots,n\}$:
\begin{align}
    r_\mu^{t+1}&=r^t_\mu-\gamma\sum_{\nu=1}^ns_\nu^t \, \ell'\Big(\bm{x}_\nu^\top\bm{w}^t,y_\nu\Big)\bm{x}_\nu^\top\bm{x}_\mu-\gamma\lambda \,r_\mu^t,\\
    \label{eq:reduced2}
    &=(1-\gamma\lambda)r^t_\mu-\gamma\sum_{\nu(\neq\mu)} s_\nu^t\, \ell'\Big(\bm{x}_\nu^\top\bm{w}^t,{y}_\nu\Big)\bm{x}_\nu^\top\bm{x}_\mu -\gamma\,s_\mu^t \ell'\Big({r}_\mu^t,{y}_\mu\Big)\bm{x}_\mu^\top\bm{x}_\mu .
\end{align}
We now consider a reference system where the first direction is parallel to $\bm{x}_\mu$ (a unit vector in the infinite-dimensional limit): then, Eq.~\eqref{eq:reduced2} describes the dynamics of the weight variable $w_1$. Notice that we have separated the term $-\gamma\,s^t\ell'\Big(r^t_\mu,y_\mu\Big)$ from the rest of the sum, to highlight that this term is of order one in the infinite-dimensional limit, at variance with the other terms, and we have used that $\bm{x}_\mu^\top\bm{x}_\mu\overset{d\rightarrow\infty}{\longrightarrow }1$. We therefore anticipate that the final equation would be formally identical to Eq.~\eqref{eq:effective_nu}, provided that this extra term is added. We report below the derivation based on the cavity method, in particular in the form introduced in \cite{liu2021dynamics}. Alternative derivations can be found in \cite{agoritsas2018out,mignacco2021stochasticity}. We proceed by solving the system of equations along all the other directions, orthogonal to $\bm{x}_\mu$, and then plugging this solution into Eq.~\eqref{eq:reduced2}, in order to obtain a self-consistent process for the effective pre-activation $r_\mu^t$. Let us denote by $\bm{\bar w}=\Big(w_2,\ldots,w_d\Big)$ the remaining directions, and similarly $\bm{\bar x}=\Big(x_2,\ldots,x_d\Big)$. Therefore, $r_\nu=\bm{\bar x}_\nu^\top\bm{\bar w}+x_{\nu,1}\,r_\mu=\bm{\bar x}_\nu^\top\bm{\bar w}+o_d(1)$, $\forall \nu\neq \mu$, since the samples are independent. We can therefore compute the solution for the dynamics of $\bm{\bar w}$ up to linear order in perturbation theory. The zeroth-order term is 
\begin{equation}
    \bm{\bar w}_0^{t+1}=(1-\gamma \lambda)\bm{\bar w}_0^{t}-\gamma\sum_{\nu\neq \mu}s_\nu^t\,\ell'\left(\bm{\bar x}_\nu^\top \bm{\bar w}_0^t\right)\bm{\bar x}_\nu.
\end{equation}
The linear-order perturbation is
\begin{align}
    \bm{\bar w}^{t}=\bm{\bar w}_0^{t}+\gamma\sum_{\nu\neq\mu}^n\sum_{t'=0}^{t-1}\frac{\delta \bm{\bar w}^t}{\delta h_\nu^{t'}}\biggr\rvert_{h_\nu=0}x_{\nu,1}\,r_\mu^{t'}, \qquad h_\nu^{t}:=x_{\nu,1}\,r_\mu^{t}.\label{eq:solution_environment}
\end{align}
We can finally plug the solution in Eq.~\eqref{eq:solution_environment} into Eq.~\eqref{eq:reduced2}. We obtain
\begin{align}
\label{eq:closed_rt}
\begin{split}
    r_\mu^{t+1}&=(1-\gamma\lambda)r_\mu^t-\gamma s_\mu^t\ell'\left(r_\mu^t,y_\mu\right)-\gamma\sum_{\nu\neq\mu}s_\nu^t\,\ell'\left(\bm{\bar x}_\nu^\top\bm{\bar w}^t_0\right){x}_{\nu,1}\\&-\gamma\sum_{\nu,\nu'\neq\mu}\sum_{t'=0}^{t-1}\frac{\delta \ell\Big(\bm{\bar x}_\nu^\top\bm{\bar w}^t_0\Big)}{\delta h_{\nu'}^{t'}}\biggr\rvert_{h_{\nu'}=0}{x}_{\nu,1}\,x_{\nu',1}\,r_\mu^{t'}-\gamma\sum_{\nu\neq\mu}s_\nu^t\ell''\left(\bm{\bar x}_\nu^\top\bm{\bar w}^t_0\right)(x_{\nu,1})^2r_\mu^t.
    \end{split}
\end{align}
We can now compute the infinite-dimensional limit of each term in Eq.~\eqref{eq:closed_rt}. Since the components of $\bm{x}_\nu$ are independent, the first sum $-\gamma \sum_{\nu\neq\mu}s_\nu^t\ell'\left(\bm{\bar x}_\nu^\top\bm{\bar w}_0^t\right)x_{\nu,1}$ reduces to a Gaussian process with zero mean and covariance
\begin{align}
    \alpha\gamma^2\mathbb{E}\left[s_\nu^ts_\nu^{t'}\ell'\left(r_\nu^t\right)\ell'\left(r_\nu^{t'}\right)\right]=C_g(t,t').
\end{align}
The term $\nu'=\nu$ dominates the second sum $-\gamma\sum_{\nu,\nu'\neq\mu}\sum_{t'=0}^{t-1}\frac{\delta \ell\Big(\bm{\bar x}_\nu^\top\bm{\bar w}^t_0\Big)}{\delta h_{\nu'}^{t'}}\biggr\rvert_{h_{\nu'}=0}{x}_{\nu,1}\,x_{\nu',1}\,r_\mu^{t'}$ that converges to
\begin{align}
    -\alpha\gamma\sum_{t=0}^{t-1}\mathbb{E}\left[s_\nu^t\frac{\delta\ell'(r_\nu^t)}{\delta h^{t'}_{\nu'}}\biggr\rvert_{h_{\nu'}=0}\right]r_\mu^{t'}= \sum_{t'=0}^{t-1}R_g(t,t')r_\mu^{t'}.
\end{align}
Similarly, the last term concentrates to
\begin{align}
    -\alpha\gamma\mathbb{E}\left[s_\nu^t\ell''\left(r_\nu^t\right)\right]r^t_\mu=\Gamma^t\,r^t_\mu.
\end{align}
Notice that the generalization performance for the problem under consideration only depends on the cosine similarity between the weight vector and the signal \cite{aubin2020generalization}. Therefore, we are interested in computing their scalar product, called \emph{magnetization} in the statistical physics literature: $m^t=\lim_{d\rightarrow\infty}\mathbb{E}\left[{\bm{w^*}}^\top\bm{w}^t\right]$, that can be obtained by multiplying both sides of the weight update Eq.~\eqref{eq:sgd_iteration} by ${\bm{w^*}}^\top$ and taking the infinite-dimensional limit. We find:
\begin{align}
    m^{t+1}=(1-\lambda\gamma)m^t-\upsilon^t,
\end{align}
where we have defined the auxiliary function:
\begin{align}
    \upsilon^t\:=\alpha\gamma\mathbb{E}\left[s^t_\nu \ell'\left(r_\nu^t\right)r_\nu^*\right], \qquad r_\nu^*=\bm{x}_\nu^\top\bm{w^*}.
\end{align}
Notice that an alternative equation for the magnetization can be found observing that \\ $ \mathbb{E}\left[{\bm{w^*}}^\top\bm{w}^t\right]\overset{d\rightarrow\infty}{\longrightarrow}\mathbb{E}\left[\theta^*\theta^t\right]$, where $\theta^t$ is drawn from Eq.~\eqref{eq:effective_nu} and $\theta^*\sim\mathcal{N}(0,1)$ is drawn from the same distribution as the signal components. Therefore, we can consider the translated variable $r_\mu^t\leftarrow r_\mu^t-r_* m^t$ and write the system of equations:
\begin{align}
    &r^{t+1}=(1-\lambda\gamma+\gamma^t)r^t-\gamma \,s^t \ell'\left(r^t+r_*m^t\right)+\sum_{k=0}^{t-1}R_g(t,k)h^k+u^t,\\
    &m^{t+1}=(1-\gamma\lambda)m^t-\upsilon^t,
\end{align}
where $u^t$ is a Gaussian process with covariance $C_g$ and we have dropped the index $\mu$ since all the samples are statistically equivalent. The above system corresponds to the one presented in Eq.~\eqref{eq:eff_eta_code} in the main text, where we have renamed $r$ by $\eta$, and that we integrate numerically.
Finally, in order to compute the cosine similarity we also need the norm of the weights as a function of time. The norm $C_\theta(t,t)=\mathbb{E}\left[(\theta^t)^2\right]$ can be computed once the convergence of the kernels has been reached, by generating multiple realizations of the stochastic process for the effective weight $\theta^t$ in Eq.~\eqref{eq:effective_nu} and computing the averages.

\bibliographystyle{siamplain}
\bibliography{references}
\end{document}